\documentclass{aa}
\usepackage{graphicx}
\usepackage{latexsym}
\usepackage{txfonts}
\usepackage{bm}

\begin{document}

\title{Detection of 20--30$\bm{h^{-1}}$Mpc-scale Galaxy Structures Embedded in 100${\bm h^{-1}}$Mpc-scale Structures of Quasars and MgII Absorbers at $\bm{z\simeq0.8}$ and $\bm{z\simeq1.2}$ \thanks{Based on observations at the European Southern Observatory, La Silla, Chile (ESO 69.A-0517), and the Cerro-Tololo Inter-American Observatory, a division of NOAO and operated by AURA under cooperative agreement with the National Science Foundation.}}
\titlerunning{Detection of 20--30$\mpc$-scale Galaxy Structures at $z\simeq0.8$ and $z\simeq1.2$}
\author{C. P. Haines\inst{1,}\inst{2,}\inst{3} 
\and L. E. Campusano\inst{2}
\and R. G. Clowes\inst{4}}

\institute{INAF - Osservatorio Astronomico di Capodimonte, v. Moiariello 16, 80131 Napoli, Italy
\and Departamento de Astronom\'{\i}a, Universidad de Chile, Casilla 36-D, Santiago, Chile
\and Centre for Astrophysics, University of Central Lancashire, Preston PR1 2HE, UK
\and Computational Astrophysics, Department of Computing, University of Central Lancashire, Preston PR1 2HE, UK}
\offprints{C. P. Haines, \email{chris@na.astro.it}}

\date{Received; Accepted}
\def\mpc {\,h^{-1}\mathrm{Mpc}}
\def\sqmpc {\,h^{-2}\mathrm{Mpc}^{2}}
\def\kpc {\,h^{-1}\mathrm{kpc}}
\def\etal  {{\em et al. }}
\def\sqarcmin {\,\mathrm{arcmin}^{2}}
\def\mgii {\ion{Mg}{ii} }
\def\gsim{ \,\lower .75ex \hbox{$\sim$} \llap{\raise .27ex \hbox{$>$}} \,}
\def\lsim{ \,\lower .75ex \hbox{$\sim$} \llap{\raise .27ex \hbox{$<$}} \,}

\abstract{We report the finding of two large-scale structures of galaxies in a \mbox{$40\times35\sqarcmin$} field embedded in a $25\deg^2$ area where two 100$\mpc$-scale structures of quasars and \mgii absorbers at $z\simeq0.8$ and $z\simeq1.2$ have been previously detected. 
Using deep optical ($V$- and $I$-band) imaging, we are able to select high-redshift \mbox{($0.5\lsim z\lsim 1.3$)} early-type galaxies as those redder than the cluster red sequence at $z=0.5$ and having $I<23.5$. 
Through comparison with a \mbox{$35\times35\sqarcmin$} control field, we find a 30\% excess of these {\em red galaxies}, corresponding to 563 galaxies across the field. The colour distribution of the excess galaxies shows a coherent peak at \mbox{$2.7<V-I<3.1$} and the magnitude distribution is well described by a Schechter function, showing unambiguously that the galaxy excess is primarily due to a population of early-type galaxies at \mbox{$z=0.83\pm0.08$}. 
In follow-up NIR imaging of the four best cluster candidates, we identify three clusters at \mbox{$z=0.8\pm0.1$}, each having well defined red sequences of \mbox{$\sim20$} galaxies at \mbox{$I-K_{s}\simeq3.4$}, forming a structure \mbox{$20\mpc$} across. 
The fourth candidate, corresponding to a previously identified cluster at \mbox{$z=1.2\pm0.1$}, is confirmed with 28 \mbox{$K_{s}<20.5$} galaxies having $VIJK_{s}$ colours consistent with passively-evolving galaxies at \mbox{$z\simeq1.2$} located within 1\,arcmin of the cluster centre. 
In all four NIR fields we find numerous galaxies with the $VIJK_{s}$ colours of passively-evolving galaxies both at \mbox{$z\simeq0.8$} and \mbox{$z\simeq1.2$}, including regions well away from the clusters.
The results of this study support the presence of two sheet-like large-scale galaxy structures extending across the 20--3$0\mpc$ scales explored by the optical data, one ``sheet'' at $z\simeq0.8$ with three associated clusters, and a second ``sheet'' at $z\simeq1.2$ with one embedded cluster. 
Our results confirm that, at least inside the probed region, the two known 100$\mpc$-scale structures at \mbox{$z\sim1$} of quasars and \mgii absorbers also mark out volumes with an enhanced density of galaxies.
The finding of two galaxy structures at \mbox{$z\simeq 1$} on scales of 20--30$\mpc$ within much larger superstructures of quasars and \mgii absorbers, firstly establishes the presence of galaxy sheet-like structures at $z\sim1$, and secondly, suggests that they may extend up to 100$\mpc$-scales.

}

\maketitle

\def\mgii {\ion{Mg}{ii} }

\section{INTRODUCTION}

Over the last twenty years galaxy redshift surveys have successfully mapped the local universe, revealing that on scales of \mbox{$\gsim100\mpc$} the spatial distribution of galaxies appears to be cellular or sponge-like, with patterns of wall-like superstructures having typical extents of \mbox{30--50$\mpc$} and thicknesses of \mbox{$\sim5\mpc$}, surrounding low-density regions or voids \mbox{50--70$\mpc$} across (Doroshkevich \etal \cite{d99}, \cite{d00}, \cite{d01}). Recent surveys covering representative volumes of the universe have established that the sheet-like large-scale structures such as the ``Great Wall'' of galaxies (Geller \& Huchra \cite{geller}), that were the most striking features of early surveys are typical phenomena (e.g. Shectman \etal \cite{shectman}; Doroshkevich \etal \cite{d96}, \cite{d00}, \cite{d02}), with average  extents of \mbox{50--100$\mpc$}, comprising $\simeq5$0\% of the galaxies with an overdensity of $\sim$5--10 above the mean (Doroshkevich \etal \cite{d00}), their distribution within these sheets being inhomogeneous with galaxies concentrated in clusters and filaments (e.g. Ramella \etal \cite{ramella}). 
There are, nevertheless, indications for deviations out to scales of \mbox{$\sim160\mpc$} (Best \cite{best}), and a turnover in the galaxy (cluster) power spectrum at these scales is still disputed (Miller \& Batuski \cite{mb}). The ``Great Wall'' appears to be representative of the upper limit in size \mbox{($\sim100\mpc$)} for these sheet-like entities belonging to the observed super large-scale structure, and as such they are expected to be rare.

The ``standard'' model for the origin and evolution of the universe is the Cold Dark Matter (CDM) model (Blumenthal \etal \cite{blumenthal}), whose most recent version has passed stringent consistency tests provided by analysis of the CMB, supernovae, and galaxy surveys (Spergel \etal \cite{spergel}). 
The $\Lambda$CDM model together with observational constraints implies that $\sim90$\% of the mass is dark, thus the distribution of luminous matter such as that of galaxies is superposed upon a dominant background DM distribution. 
Our ignorance of the relation between specific luminous markers (i.e. intergalactic gas, galaxies, radio galaxies, quasars) and the underlying DM distribution is expressed through a ``bias'' parameter or function (Kaiser \cite{kaiser}; Blanton \etal \cite{blanton99}), which represents one of the most fundamental sources of uncertainty in the efforts to perform meaningful comparisons between the observations of the galaxy distribution and the DM model predictions. 

With these caveats in mind, the observed large-scale structure in the galaxy distribution may be understood in the context of the Zel'dovich non-linear theory of gravitational instability, whereby perturbations in the initial Gaussian random density field collapse rapidly along one axis to form Zel'dovich pancakes (sheets), with further collapses along the remaining axes to form filaments and clusters (Demianski \& Doroshkevich \cite{dd99a}, \cite{dd99b}). 
Numerical simulations have allowed the evolution of large-scale structure from the initial Gaussian perturbations to be followed up to the present for a variety of CDM models, successfully reproducing the main properties of the observed large-scale galaxy distribution (e.g. Doroshkevich \etal \cite{d99}). 
The most stringent constraints on cosmological models should be provided by direct observation of this evolution through measuring the properties of large-scale structures at the highest possible redshifts; a task beyond the scope of the current large galaxy redshift surveys which reach only \mbox{$z\sim0$.2--0.3}. 
One approach to investigate large-scale structure at higher redshifts is to use quasars, which are much easier to detect at \mbox{$z>1$} than galaxies. At low-redshifts \mbox{($z<0.3$)}, groups of quasars and AGN are shown to delineate the underlying large-scale structure of galaxies and galaxy clusters (e.g. S\"{o}chting \etal \cite{sochting}), a particular example being the ``Great Wall'' which is traced by a group of 19 AGN (Longo \cite{longo}). At higher redshifts \mbox{($z\gsim0.5$)} a number of quasar superstructures have been discovered, made up of 10--25 quasars, and spanning \mbox{50--20$0\mpc$} (e.g. Webster \cite{webster}; Crampton \etal \cite{crampton87}, \cite{crampton89}; Clowes \& Campusano \cite{clowes91}, \cite{clowes94}). These large quasar groups (LQGs), if proved to be marking out volumes with a generalised galaxy overdensity, may represent the precursors of the ``Great-Wall''-like structures such as those found in the local universe (Komberg \& Lukash \cite{komberg}), and hence may provide ideal laboratories for studying the evolution of super-large scale structure of galaxies.

The relationship between the quasar distribution and the underlying galaxy or mass distribution is likely to be both complex and redshift-dependent, particularly on cluster length-scales \mbox{($\lsim2\mpc$)} where the dependence on the environment of the fuelling mechanism of the quasar is likely to be dominant. On larger scales, the quasar spatial distribution should mirror that of their host galaxies, the majority of which, from high-resolution observations of bright quasars, appear to be massive ellipticals (e.g. McLure \etal \cite{mclure99}; Kukula \etal \cite{kukula}), which are known to be already in place at \mbox{$z\sim1.2$} (Blakeslee \etal \cite{blakeslee}). This result is consistent with the requirement of a massive black-hole to power the quasar, in conjunction with the observed correlation between the masses of the black-hole and the bulge of the host galaxy (Merritt \& Ferrarese \cite{merritt}). The expectation that quasars and their host galaxies (i.e. massive ellipticals) trace the underlying mass distribution in similar ways also appears confirmed through comparison of the quasar and galaxy power-spectra (Hoyle \etal \cite{hoyle}) and 2-point correlation-functions (Croom \etal \cite{croom}) which show the same general form over the range \mbox{2--3$0\mpc$}.

\begin{figure*}[t]
\centerline{{\resizebox{17cm}{!}{\includegraphics{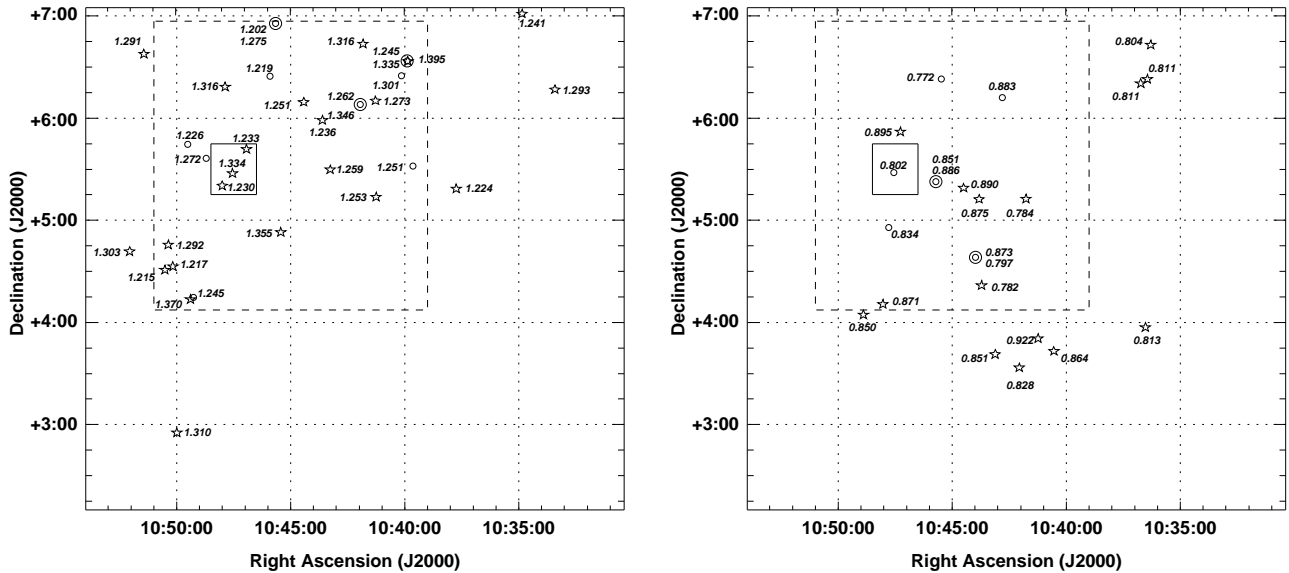}}}}
\caption{The Clowes-Campusano LQG {(\bf left; \mbox{$\bm{1.2<z<1.4}$})} and the $z\simeq0.8$ structure {(\bf right; \mbox{$\bm{0.77<z<0.93}$})}. Each quasar and \mgii absorber within the redshift range of the two structures are indicated by stars and circles respectively, labelled by their redshift. Double circles indicate two \mgii systems towards a line of sight. The boundaries of the plots match the boundaries of the AQD survey of ESO/SERC field 927 (Clowes \& Campusano \cite{clowes91}, \cite{clowes94}), while the boundaries of the \mgii survey (Williger \etal \cite{williger}) and the BTC field (Haines \cite{hainesphd}) are indicated by dashed and solid boxes.}
\label{lqgs}
\end{figure*}

Of the quasar groups identified to date, the largest spatially is that of the Clowes-Campusano LQG, with at least 18 quasars at \mbox{$1.2<z<1.4$} towards ESO/SERC field 927 (Clowes \& Campusano \cite{clowes91}, \cite{clowes94}; Graham \etal \cite{graham95};  Newman \etal \cite{newman98}; Clowes \etal \cite{clowes99}), with a spatial overdensity of 6--10 times over the mean. The structure spans \mbox{$\sim2.5\degr\times5\degr$} on the sky, corresponding to \mbox{$\sim120\times240\sqmpc$} at \mbox{$z\simeq1.3$}, making it the largest known structure in the high redshift universe.

To investigate the underlying galaxy distribution traced by the Clowes-Campusano LQG, a study of \mgii absorption systems in the spectra of background quasars has been carried out (Williger \etal \cite{williger}). Spectra for 23 quasars \mbox{($1.23<z<2.68$)} in a \mbox{$2.5\degr\times2.5\degr$} field towards the centre of the LQG were obtained. The advantage of using \mgii absorbers is that they trace much lower mass overdensities than quasars, and thus offer a much more detailed picture of the overall mass distribution.
 38 \mgii absorbers were identified \mbox{($W_{0}>0.3$\AA)} while only 24 would be expected given the quasars observed and the wavelength coverage of the survey. Of these, 11 were found to be associated with the LQG having \mbox{$1.2<z<1.4$} whereas only \mbox{$3.8\pm1.8$} would be expected, a result significant at the \mbox{1--2\%} level and confirming the hypothesis that the quasar group delineates an underlying galaxy large-scale structure. The \mgii absorber survey also identified a second structure at \mbox{$z\simeq0.8$} made up of 7 \mgii absorption systems at \mbox{$0.77<z<0.89$}, again significant at the \mbox{1--2\%} level. This structure coincides with a group of 14 quasars at \mbox{$z\sim0.8$} previously identified by the Chile-UK Quasar Survey (CUQS, Newman \cite{newman99}) as a structure significant at the 6.6\% level. This structure spans \mbox{$\sim3.5\degr\times3\degr$} on the sky, corresponding to \mbox{$\sim120\times100\sqmpc$} at \mbox{$z\simeq0.8$}.

Figure~\ref{lqgs} shows the Clowes-Campusano LQG (left; \mbox{$1.2<z<1.4$}) and the \mbox{$z\simeq0.8$} structure (right; \mbox{$0.77<z<0.93$}). Each quasar and \mgii absorber within the redshift range of the two structures are indicated by stars and circles respectively, labelled by their redshift. The boundaries of the plots match the boundaries of the AQD survey of ESO/SERC field 927 (Clowes \& Campusano \cite{clowes91}, \cite{clowes94}), while the boundaries of the \mgii survey (Williger \etal \cite{williger}) and the BTC field (a \mbox{$40\times35\sqarcmin$} region covered by deep $V$ and $I$ imaging; Haines \cite{hainesphd}) are indicated by dashed and solid boxes respectively. The detected \mgii absorbers at \mbox{$z\simeq0.8$} and \mbox{$z\simeq1.2$} are found in the same regions as the large quasar concentrations detected at the same redshifts.  

To directly detect part of the underlying galaxy distribution accompanying the large-scale structures observed in the quasar and \mgii absorber distributions at \mbox{$z\simeq0.8$} and \mbox{$z\simeq1.2$}, we have also undertaken an ultra-deep optical imaging survey of a \mbox{$40\times35\sqarcmin$} region -- the BTC field -- containing three quasars from the Clowes-Campusano LQG, in both $V$ and $I$ passbands. These data were used previously by Haines \etal (\cite{haines01a}) in conjunction with UKIRT $K$-band imaging to identify a cluster at \mbox{$z\simeq1.2$} associated with the \mbox{$z=1.233$} LQG quasar (redshift value changed from \mbox{$z=1.226$} as reported in Haines \etal \cite{haines01a} based on new spectroscopic data; Williger \etal \cite{williger}). A factor $\sim$11 overdensity of \mbox{$I-K>3.75$} galaxies was identified in a \mbox{$2.25\times2.25\sqarcmin$} field centred on the \mbox{$z=1.233$} LQG quasar, for which NIR imaging had been obtained, reaching \mbox{$K\simeq20$}. In particular, 15--18 galaxies with colours consistent with being a population of passively-evolving massive ellipticals at the quasar redshift were found. This field was reobserved using the New Technology Telescope (NTT) with its infrared camera (SOFI) for the study described in this article, and is part of field 1 (see Sect. 5). At the redshift of the Clowes-Campusano LQG ($z=1.2$) the BTC field covers $31\times27\sqmpc$, while at $z=0.8$ the field covers \mbox{$23\times20\sqmpc$}. These observations along with follow-up NIR imaging are described in Sect. 2.

To identify any possible clustering or large-scale structure in the form of galaxies at $z\sim1$ a variant of the cluster red sequence (CRS) algorithm of Gladders \& Yee (\cite{gladders}) is applied to the BTC dataset. This approach is motivated by the observation that the bulk of early-type galaxies in clusters lie along a tight, linear colour-magnitude relation, the CRS, that becomes increasingly red with redshift (depending on choice of filters). In Sect. 3 we describe the variant of the CRS algorithm used, whereby galaxies redder than the \mbox{$z=0.5$} CRS are selected, and demonstrate its ability to identify early-type galaxies at $0.5\lsim z\lsim 1.3$, i.e. the redshift range of interest. In Sect. 4 the number density of these {\em red galaxies} in the BTC field is compared with that of a similar control field (in terms of passbands, depth and area covered) taken from the Deep Lens Survey (DLS) of Wittman \etal (\cite{wittman}). Evidence is found for there being one or more large-scale structures at \mbox{$z\sim1$} in the BTC field manifested as an overall excess of red galaxies. In Sect. 5 we describe follow-up targeted deep near-infrared imaging of four \mbox{$z\sim1$} cluster candidates identified as density enhancements in the red galaxy spatial distribution which also have apparent red sequences at \mbox{$V-I\simeq2.8$}. The implications of our results are discussed in Sect. 6 and the conclusions presented in Sect. 7.

Throughout the paper we assume a flat, $\Lambda$-dominated cosmology with \mbox{$\Omega_{m}=0.3$}, \mbox{$\Omega_{\Lambda}=0.7$}, and for the predictions of the evolution of galaxy colours we assume \mbox{H$_{0}=70\,\mathrm{km\,s}^{-1}\,\mathrm{Mpc}^{-1}$}. With these parameters, the age of the Universe is 13.5\,Gyr, and the redshifts \mbox{$z=0.8$} and \mbox{$z=1.2$} correspond to look-back times of $6.8$ and $8.4$\,Gyr respectively.

\section{OBSERVATIONS}

\subsection{Optical imaging data}

The optical data were obtained using the Big Throughput Camera (BTC) on the 4-m Blanco telescope at the Cerro Tololo Inter-American Observatory on April 21/22 and 22/23 1998. The resulting BTC field (as we shall refer to it) covers \mbox{$40\times35\sqarcmin$} in both $V$ and $I$ imaging, reaching \mbox{$V\simeq26.35$} and \mbox{$I\simeq25.85$}. It is centred at \mbox{$10^{h}47^{m}30^{s}$}, \mbox{$+05\degr32\arcmin00\arcsec$} (J2000) and was selected to contain three members of the Clowes-Campusano LQG.

The BTC is made up of four \mbox{$2048\times2048$} CCDs which have pixels of size 0.43\,arcsec, giving a field of view for each CCD of \mbox{$14.7\times14.7\sqarcmin$}. The CCDs are arranged in a 2 by 2 grid and are separated by gaps of 5.6\,arcmin. To obtain a contiguous image the camera had to be shifted between exposures. The camera geometry meant that it was not possible to obtain a uniform coverage of the field. This, along with one of the CCDs being markedly less sensitive (55\% less in $V$, 30\% in $I$) than the other three, meant that there were significant variations in depth across the field, and so separate magnitude limits have been determined for each subregion discussed in this paper. Further details of the observing and reduction procedures for the BTC images are presented elsewhere (Haines \cite{hainesphd}).

\subsection{Near-infrared imaging data}

Deep near-infrared imaging data were obtained using the SOFI camera on the 3.5-m ESO New Technology Telescope (NTT) on April 1--2 2002. The SOFI camera has a Rockwell HgCdTe \mbox{$1024\times1024$} Hawaii array with a pixel size of 0.292\,arcsec giving a field of view of \mbox{$4.9\times4.9\sqarcmin$} which corresponds to comoving angular scales of \mbox{$3.8\times3.8\sqmpc$} at \mbox{$z=1.2$} and \mbox{$2.8\times2.8\sqmpc$} at \mbox{$z=0.8$}. Given the field of view, targeted fields within the BTC field were selected as having high-densities of red galaxies and apparent red sequences at \mbox{$V-I\simeq2.8$} that may indicate galaxy clusters at \mbox{$z\sim1$}. The best three candidates were observed for 3600s through both $J$ and $K_{s}$ filters reaching \mbox{$J\simeq23$} and \mbox{$K_{s}\simeq21$}, and another four fields were observed for 900s in $K_{s}$ reaching \mbox{$K_{s}\simeq20$}. It was found after the first night that the sky background levels in the $K_{s}$-band fell by a factor of three during the first half of the night, and so field 1, which had been observed at the start of night one in $K_{s}$, was reobserved for a further 3600s on the second night in more favourable conditions. Conditions were photometric throughout and the seeing was 0.9--1.5\,arcsec. 

The NIR imaging data were reduced using standard IRAF routines. To allow the estimation of sky levels to be made from the observations themselves, each integration of 60s duration, and made up of 3(6) sub-integrations for the $J(K_{s})$-bands, was jittered randomly in a 20\,arcsec wide box. For each individual exposure, the sky background was estimated by combining the 10 integrations closest in time by the median, after rejecting all pixel values greater than $3\,\sigma$ from the median. The sky image is then subtracted from its corresponding image after being scaled to have the same median value. The images are then flat-fielded through the use of dome flats taken using the {\em SOFI\_ima\_cal\_SpecialDomeFlat} template which allows a residual shade pattern to be estimated and removed. An illumination correction was then applied to the images using illumination correction surfaces produced by observing a photometric standard star at differing parts of the array and fitting a surface. The SOFI array suffers from interquadrant row cross talk, which produces ghosts that affect all the rows of a bright source, and also the corresponding rows in the opposing half of the detector. Although the effect is not completely understood, it is well described and was corrected for in each of the raw exposures. The individual exposures were then registered and coadded to produce the final images.

\section{SELECTION OF EARLY-TYPE GALAXIES AT $\bm{0.5<z<1.3}$}

To detect and identify any possible large-scale structure at \mbox{$z\sim1$} a variant of the cluster red sequence algorithm of Gladders \& Yee (\cite{gladders}) is applied to the BTC dataset. This approach is motivated by the observation that the bulk of early-type galaxies in all rich clusters lie along a tight, linear colour-magnitude (C-M) relation -- the cluster red sequence. Studies of low-redshift clusters (e.g. Bower \etal \cite{bower}) indicate that, irrespective of the richness or morphology of the cluster, all clusters have red sequences, whose k-corrected slopes, scatters and colours are indistinguishable. This indicates that the early-type galaxies which make up the red sequences form a homogeneous population, not only within each cluster, but from cluster to cluster, and also that red sequences are universal and homogeneous features of galaxy clusters, at least at \mbox{$z\lsim 0.2$}. The photometric evolution of the cluster red sequence with redshift has been studied (e.g. Arag\'{o}n-Salamanca \etal \cite{aragon}; Ellis \etal \cite{ellis}; Stanford \etal \cite{stanford98}; Kodama \etal \cite{kodama}) for clusters out to \mbox{$z\simeq1.2$} indicating: that red sequence remains a universal feature of clusters; that the stellar populations of its constituent early-type galaxies are formed in a single, short burst at an early epoch \mbox{($z_{f}\gsim 3$)}; and that the galaxies have evolved passively ever since. This is confirmed by the spectra of red-sequence galaxies which in nearby clusters are best fit by simple stellar populations of ages \mbox{9--12\,Gyr}, resulting in their having characteristically strong spectral breaks at 4000\,\AA, and correspondingly red $U-V$ colours. 

The basic implementation of the cluster red sequence algorithm is to split the galaxy catalogue into a series of redshift {\em slices} by selecting the subset of galaxies whose colours and magnitudes are consistent with the C-M relation of clusters at that redshift. For each slice the galaxy surface density is then estimated, and significant peaks in each slice are then identified as cluster candidates at that redshift. The key advantages of this algorithm for cluster detection are: that it requires only photometry in two passbands, as is the case here; the colour of the observed red sequence can be used as a precise redshift indicator; and it does not suffer from projection effects or foreground contamination. The latter two are contingent on the passbands used covering the rest-frame 4000\AA\, break for the redshift range of interest, as then the evolution of colour of the red sequence with redshift is greatest allowing the most accurate results, and also the red sequence galaxies are as red or redder than all galaxies at that redshift and all nearer galaxies, hence eradicating the problem of foreground contamination. Hence for our data with $V$ and $I$ photometry, the optimal redshift range is \mbox{$0.3\lsim z\lsim 0.8$}, and conversely the optimal passbands for estimating redshifts of clusters at \mbox{$0.8\lsim z\lsim 1.3$} are $I$ and $K_{s}$.

\begin{figure}[t]
\centerline{{\resizebox{\hsize}{!}{\includegraphics{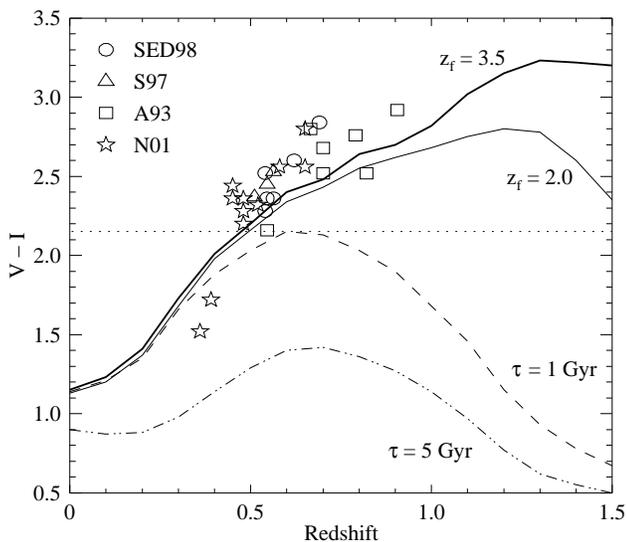}}}}
\caption{Evolution of galaxies' $V-I$ colours as a function of galaxy type and redshift. The solid curves represent stellar populations formed in an instantaneous burst at $z_{f}=3.5$ (thick) and $z_{f}=2.0$ (thin), and are thought to represent the colour evolution of massive ellipticals. The dashed and dot-dashed curves represent stellar populations with exponentially-decaying star-formation rates with time-scales ($\tau$) of 1 and 5\,Gyr respectively, and are thought to represent the colour evolution of disk-dominated galaxies. The points correspond to the colour of the red sequence for clusters from the surveys of SED98, S97, A93 and N01.
The dotted line indicates the colour selection threshold for red galaxies at $I=23.5$.}
\label{modelcols}
\end{figure}

Figure~\ref{modelcols} shows the evolution of galaxies' $V-I$ colours as a function of galaxy type and redshift. The solid curves indicate the expected evolution of the red sequence with redshift, the most massive red sequence galaxies being more likely to form early ($z_{f}\simeq3.5$) as shown by the thick line, while the less massive early-type galaxies are more likely to follow the thin line, indicating later formation epochs ($z_{f}\simeq2.0$) or lower metallicities. The dashed and dot-dashed curves correspond to stellar populations with exponentially-decaying star-formation rates with time-scales ($\tau$) of 1 and 5\,Gyr respectively, and are thought to represent the colour evolution of disk-dominated galaxies. The points correspond to the $V-I$ colour of the red sequence for clusters from the surveys of Stanford \etal \cite{stanford98} (SED98 : 6 clusters at \mbox{$0.54<z<0.7$}), Smail \etal \cite{smail97} (S97 : 3 clusters at \mbox{$z\sim0.54$}), Arag\'{o}n-Salamanca \etal \cite{aragon} (A93 : 10 clusters at \mbox{$0.5<z<0.9$}) and Nelson \etal \cite{nelson} (N01 : 30 clusters at \mbox{$0.3<z<0.9$}). 
The figure indicates that at redshifts less than \mbox{$z\sim0.8$} the $V-I$ colour of the red sequence increases monotonically with redshift, and so can be used to efficiently estimate the cluster redshift. At higher redshifts the $V-I$ colour becomes a less efficient redshift indicator, as its dependence on redshift decreases, and dependence on star-formation history increases. The $V-I$ colour of the red sequence in clusters at \mbox{$0.8\lsim z\lsim 1.3$} remains roughly constant at \mbox{$V-I\simeq2.8$}, as the 4000\AA\, break passes through the $I$-band. However the two models used to describe the red sequence galaxies show significant divergence at \mbox{$z>1$}, indicating that the colour of red sequences in clusters at \mbox{$0.8<z<1.3$} could be in the range \mbox{$2.5<V-I<3.2$}, and so it is not possible to obtain an accurate redshift estimate based on the \mbox{$V-I$} colour of the red sequence for \mbox{$z\sim1$} clusters. At these redshifts the $V$-band now corresponds to a rest-frame wavelength of 2500--3000\AA, so the $V-I$ colour is affected by even small amounts of recent star-formation as is indicated by the divergence of the exponentially-decaying star-formation rate models from the burst models. A third problem with the basic cluster red sequence algorithm at these redshifts is that at \mbox{$z\gsim 1$} even $L^{*}$ galaxies are reaching the detection limit in $V$, and so that the red sequence is likely to be smeared in the colour direction by the less accurate photometry, reducing the signal of any cluster.

Given all these problems facing the standard cluster red sequence algorithm, it was decided that instead of considering narrow redshift slices over the redshift range \mbox{$0.8\lsim z\lsim 1.3$}, a single high-redshift slice would be used to identify structures at this redshift range, containing all galaxies redder in $V-I$ than the cluster red sequence at $z=0.5$. Kodama \etal (\cite{kodama}) show that the Kodama \& Arimoto (\cite{ka97}) evolutionary model for elliptical galaxies with \mbox{$z_{f}=4.5$} describes well the evolution of the zero-point and slope of the cluster red sequence over \mbox{$0.3<z<1.2$}. We take directly from Figs. 4 and 5 of Kodama \etal (\cite{kodama}) their values for the zero-point and slope of the cluster red sequence at \mbox{$z=0.5$}, resulting in the selection criterion for {\em red galaxies} of
\begin{equation}
\begin{array}{l}
 V-I < 3.689 - 0.0652 \times I_{total};\hspace{0.6cm} I_{total}<23.5.
\end{array}
\end{equation}
A magnitude limit is also applied, as most red sequence galaxies should be brighter than $I=23.5$ to \mbox{$z\simeq1.3$} (an $L^{*}$ early-type galaxy has \mbox{$I\simeq23.0$} at \mbox{$z=1.3$}), and any galaxy with a sufficiently red \mbox{$V-I$} colour will be close to the $V$-band detection limit. Hence all galaxies classed as belonging to the high-redshift slice will have \mbox{$V-I>2.15$}, as indicated in Fig.~\ref{modelcols} by the horizontal dotted line. An examination of this figure indicates that this approach will remove all foreground galaxies (\mbox{$z<0.5$}), as well as any star-forming galaxies, leaving only passively-evolving galaxies in the redshift range of \mbox{$0.5\lsim z\lsim 1.3$}. It should then be possible to identify any galaxy clusters or large-scale structures at this redshift range as density enhancements in the spatial distribution of these {\em red galaxies} (as we shall refer to them), but there will be little ability to distinguish across this redshift range.

\subsection{Red Galaxies in the ESO Imaging Survey of the Hubble Deep Field South}

As a test of the efficiency of the colour-magnitude criteria to select early-type galaxies at \mbox{$0.5\lsim z\lsim 1.3$}, we have applied the same criteria to galaxies in the ESO Imaging Survey (EIS) of the Hubble Deep Field South. This comprises deep optical data for a \mbox{25\,arcmin$^{2}$} field reaching $2\sigma$ limiting magnitudes of $U\sim26$, $B\sim26.5$, $V\sim26$, $R\sim26$, $I\sim24.5$ obtained using the SUSI2 imager on the NTT, plus near-infrared data reaching $J\sim24$, $H\sim22.5$ and $K_{s}\sim22$ for about 40\% of the optical field. In total 34 red galaxies (those satisfying Eq. 1) are identified in the EIS-DEEP field, of which 9 have near-infrared imaging.

\begin{figure}[t]
\centerline{{\resizebox{\hsize}{!}{\includegraphics{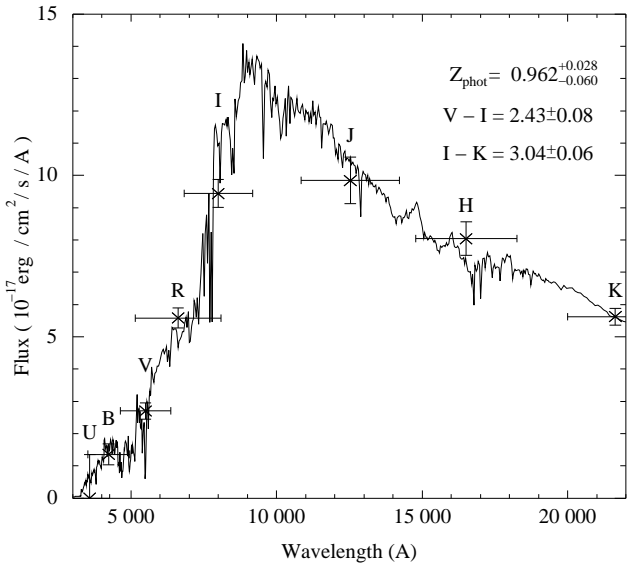}}}}
\caption{Best-fit synthetic SED to sample galaxy in EIS-DEEP field with $UBVRIJK_{s}$ photometry}
\label{eis}
\end{figure}

To estimate the photometric redshifts of each of the galaxies from its $UBVRI$ (and where possible $JHK_{s}$) photometry, and to produce the model colours tracks used throughout this paper, the {\sc hyperz} code of Bolzonella \etal (\cite{bolzonella}) has been used. It builds synthetic template galaxies using the Bruzual \& Charlot evolutionary code (GISSEL98; Bruzual \& Charlot \cite{bruzual}). It has stellar populations with eight star-formation histories roughly matching the observed properties of local galaxies from E to Im type: an instantaneous burst; a constant star-formation rate; and six exponentially-decaying star-formation rates with time-scales, $\tau$, from 1 to 30\,Gyr chosen to match the sequence of colours from E-S0 to Sd galaxies. The models assume solar metallicity and a Miller \& Scalo (\cite{miller}) initial mass function with upper and lower mass-limits for star-formation of 125\,M$_{\odot}$ and 0.1\,M$_{\odot}$ respectively. Internal reddening is also considered through the Calzetti \etal (\cite{calzetti}) model with $A_{V}$ allowed to vary between 0 and 1\,mag. The {\sc hyperz} software then produces a photometric redshift probability distribution for each galaxy through a chi-squared minimization process, allowing for all galaxy ages, star-formation histories and $A_{V}$s. Hence instead of a single best-fitting redshift for a galaxy, a range of compatible redshifts is produced.

We find that, despite allowing for all galaxy types and ages, all the red galaxies are best-fit by early-type galaxies whose stellar populations formed in an instantaneous burst. As predicted, the selection of galaxies redder in $V-I$ than the \mbox{$z=0.5$} cluster red sequence removed all foreground contamination, 32 of the 34 red galaxies in the EIS-DEEP field are constrained at the $1\sigma$ level to be at \mbox{$z>0.6$}, the other two being constrained to \mbox{$z>0.45$}. The selection of red galaxies effectively identifies a population of passively-evolving early-type galaxies at \mbox{$0.5\lsim z\lsim 1.3$}, whose strong 4000\AA\, breaks are responsible for their red $V-I$ colour. This is demonstrated by Fig.~\ref{eis} which shows the best-fitting synthetic SED to a sample galaxy (\mbox{$\alpha$=22:32:42.31}, \mbox{$\delta$=-60:33:43.0)} in the EIS-DEEP field with $UBVRIJK_{s}$ photometry. The redshift estimate is constrained to a narrow range with \mbox{$z_{best}=0.962^{+0.028}_{-0.060}$}, and it is clear that this is due to the very strong 4000\AA\, break which is manifested as the red \mbox{$V-I$} colour.

\section{RESULTS : THE BTC FIELD}

\subsection{Red galaxies in the BTC field}

To identify the red galaxies in the BTC field, catalogues of sources in the registered $V$ and $I$ images were created, using the $I$ image to detect the sources given the red colour of the sources of interest. Registration and astrometric calibration of the images was performed using the USNO astrometric catalogue, and is accurate to $\simeq0.05\,$arcsec. Photometric calibration of the $V$ and $I$ images onto the Johnson-Kron-Cousins system was obtained using 15 Landolt (\cite{landolt}) standard stars observed at varying airmasses, resulting in zero-point uncertainties of 0.026 ($V$) and 0.009 ($I$), producing an uncertainty in \mbox{$V-I$} of 0.028\,mag. The V-band uncertainty is greater due to the large differences in sensitivity across the four CCDs. Object detection was performed on the $I$-band image using {\sc SExtractor} (Bertin \& Arnouts \cite{bertin}) for objects with 4 contiguous pixels 0.8$\sigma$ over the background level. The total $I$ magnitude was taken to be the MAG\_BEST output from {\sc SExtractor} and the \mbox{$V-I$} colour determined using fixed apertures of 2.5\,arcsec diameter with {\sc SExtractor} in two image mode. 

All sources satisfying the red galaxy selection criteria {(\bf Eq. 1)} were then visually verified in the $I$ image, and spurious objects removed. The bulk of these contaminant sources are due to random noise events near the field edges, and given the faint magnitude limit of \mbox{$I=23.5$} is 2\,mags above the detection threshold of the $I$-band images, it is relatively easy to discriminate between real and spurious sources in the vast majority of cases. Regions around bright, heavily-saturated stars were also rejected as in the $I$ image the wide wings of the point-spread-functions artificially boost the $I$-band flux of faint sources, resulting in rings of ``red'' sources. In total \mbox{$\sim5\%$} of the area of the BTC field is discarded in this manner, but as stars should be randomly distributed this should not affect the results. 

The most significant source of contamination of the red galaxy subset is that from faint red dwarfs. To minimise this contamination, the in-built star-galaxy classifier of {\sc SExtractor} was applied, and 467 sources with \mbox{$\mathrm{CLASS\_STAR}>0.95$} taken to be stars and rejected. The stellarity classifier is believed to be efficient to \mbox{$I\sim22$} where the vast majority \mbox{($>90$\%)} of sources have stellarities greater than 0.95 (i.e. stars) or less than 0.2 and are hence galaxies. At fainter magnitude levels there is likely to remain some contamination due to stellar sources, as the classifier becomes less efficient, and although at this level there are two clear loci in the CLASS\_STAR distribution, presumably corresponding to sources well-classified as stars or galaxies, there remain \mbox{$\sim250$} sources with ambiguous stellarities \mbox{($0.2<\mathrm{CLASS\_STAR}<0.95$)}. 

After removing spurious and stellar sources, and discarding the regions whose photometry is affected by bright stars, 2674 red galaxies remain in a total area of \mbox{1277\,arcmin$^{2}$}. This gives a density of \mbox{$2.09\pm0.14$} galaxies arcmin$^{-2}$, where the uncertainty considers Poisson noise as well as the effect of the two-point angular correlation function $\omega(\theta)$ through the equation
\begin{equation}
\sigma^{2}=\bar{n} + \bar{n}^{2} \frac{\int\int \omega({\theta_{12}}) d\Omega_{1}d\Omega_{2}}{\int\int d\Omega_{1}d\Omega_{2}},
\end{equation}
where $\bar{n}$ is the number of galaxies, and \mbox{$\omega({\theta_{12}})=\mathrm{A}_{\omega}^\mathrm{red}(1')\,\theta^{-1.045}$} is the two-point angular correlation function of red galaxies in the BTC field (Haines \cite{hainesphd}). The red galaxies are much more strongly clustered than galaxies selected by magnitude only (i.e. \mbox{$I<23.5$}), with \mbox{A$^\mathrm{red}_{\omega}(1')=0.0863$} in comparison to \mbox{A$^{I<23.5}_{\omega}(1')=0.0175$}. As a consequence, the effect of the clustering on the rms uncertainty in galaxy counts is significant, three times that of Poisson noise alone.

\subsection{An excess of red galaxies -- A large-scale structure at $z\sim1$ across the BTC field}

Any large-scale structure at \mbox{$z\sim1$} underlying either the Clowes-Campusano LQG at \mbox{$z\simeq1.2$} or the structure of quasars and Mg{\sc ii} absorbers at \mbox{$z\simeq0.8$} should manifest itself as an excess of red galaxies across the whole BTC field. To determine the extent of any excess we estimate the expected field density of red galaxies through consideration of a suitable random control field. For this purpose we have used field F1p22 from the Deep Lens Survey (DLS) of Wittman \etal (\cite{wittman}; http://dls.bell-labs.com). The Deep Lens Survey is an ongoing ultra-deep multi-band optical survey planned to cover seven randomly-chosen 4\,deg$^{2}$ fields with deep $BVRz$ imaging obtained using the Mosaic cameras on the 4-m telescopes at the Cerro Tololo and Kitt Peak observatories. The publicly-available field F1p22 has $BVRIz$ imaging covering \mbox{$35\times35\,{\rm arcmin}^{2}$} centred at \mbox{$0^\mathrm{h}53^\mathrm{m}25\fs 30$}, \mbox{$+12\degr33\arcmin55\farcs0$} and has limiting magnitudes of \mbox{$I\sim25$} and \mbox{$V\sim26$}. It is an excellent comparison field, covering a similar size, and reaching similar depths in the same filters as the BTC field.

To identify the red galaxies in the DLS data, exactly the same detection and source extraction processes are applied as for the BTC field, with objects detected in the $I$ image using {\sc SExtractor} and colours determined through 2.5\,arcsec diameter apertures using {\sc SExtractor} in two-image mode. As for the BTC field, red galaxies are identified as sources redder in \mbox{$V-I$} than the \mbox{$z=0.5$} cluster red sequence and brighter than $I=23.5$. Once again, regions around bright stars are discarded, spurious objects are rejected through visual inspection, and stars identified by the star-galaxy classifier as sources with \mbox{$\mathrm{CLASS\_STAR}>0.95$} and removed from the final catalogue. We expect the efficiency of the identification and removal of both stars and spurious objects from the red galaxy catalogues to be similar for the BTC and DLS data, as both have comparable magnitude limits \mbox{($I\approx25.5$)} and levels of seeing \mbox{($\approx1.2\,$arcsec)}, resulting in near-identical image qualities.

In total 1953 red galaxies are found in the \mbox{1181\,arcmin$^{2}$} area of the DLS image not affected by bright stars, corresponding to a mean density of \mbox{$1.65\pm0.12$ gals arcmin$^{-2}$}.

\begin{figure}[t]
\centerline{{\resizebox{\hsize}{!}{\includegraphics{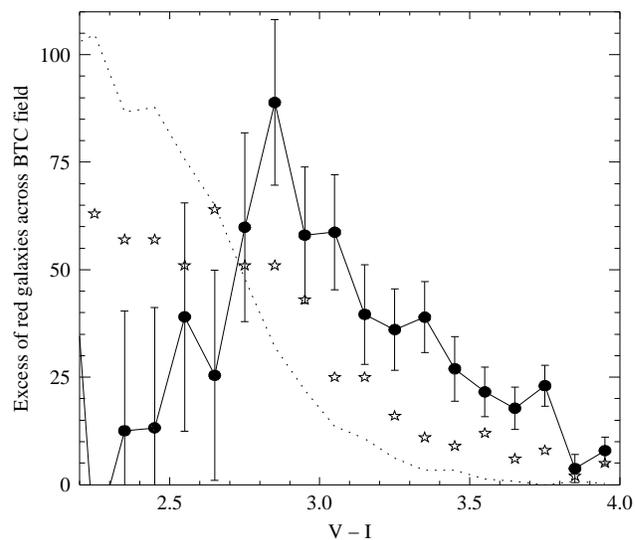}}}}
\caption{The $V-I$ colour distributions of the observed excess of red galaxies for the BTC field (solid curve connecting filled circles). For comparison the dotted curve shows the colour distribution of galaxies in the DLS field (scaled to fit plot), and the star-shaped symbols indicate the colour distribution of stars in the BTC field.}
\label{redxs}
\end{figure}

We would thus expect \mbox{$2111\pm151$} galaxies in the BTC field, where the uncertainty assumes both Poisson noise and the effect of the two-point angular correlation function as before. The density of red galaxies in the BTC field is 27\% greater than that observed for the DLS image, corresponding to an excess of 563 galaxies over the \mbox{$40\times35\,$arcmin$^{2}$} field, a result significant at the \mbox{3.7$\,\sigma$} level. 

To ascertain the nature of this excess we plot in Fig.~\ref{redxs} the  $V-I$ colour distribution of the galaxy excess in 0.1\,mag wide bins in $V-I$ as filled circles connected by a solid line. This is determined by scaling the colour distribution of galaxies in the DLS to have the same effective area as the BTC field, and subtracting it from the colour distribution of galaxies in the BTC field. The errors shown are determined as \mbox{$\sigma^{2}_{i}=N_{i}\mathrm{(BTC)}+N_{i}\mathrm{(DLS)}$}.  Significant excesses are apparent for all bins in the range \mbox{$2.5<V-I<3.8$}. The excess shows a coherent peak at \mbox{$2.7<V-I<3.1$}, and is greatest for the \mbox{$2.8<V-I<2.9$} bin for which 224 such galaxies are identified in the BTC field whereas only 135 would be expected given the density in the DLS image, corresponding to a 66\% excess, significant at the \mbox{3.5$\,\sigma$} level. A comparison with Fig.~\ref{modelcols} shows that this peak corresponds to the colours expected of passively-evolving early-type galaxies at the redshift range of interest, \mbox{$0.8\lsim z\lsim 1.3$}. 
For comparison the star symbols indicate the colour distribution of stars (defined to have \mbox{$\mathrm{CLASS\_STAR}>0.95$)} in the BTC field. The colour distributions of stars and the excess red galaxies appear quite different, indicating that stellar contamination is not a major contributor to the red galaxy excess. The colour distribution of galaxies in the DLS field (scaled to fit plot) is also shown by the dotted curve, and again appears different to the distribution of the excess red galaxies, indicating that the excess forms a population that is quite distinct to that from the field.

\begin{figure}[t]
\centerline{{\resizebox{\hsize}{!}{\includegraphics{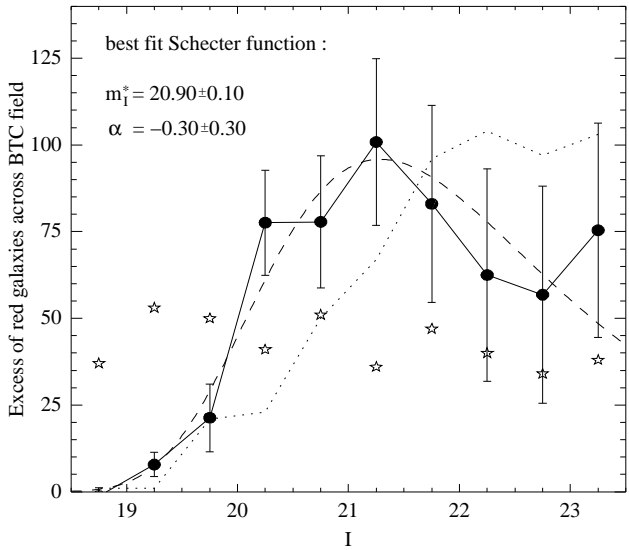}}}}
\caption{The $I$ magnitude distribution of the observed excess of red galaxies across the BTC field shown as a solid curve connecting filled circles. The dashed curve corresponds to the best-fitting Schechter function with $\alpha=-0.3$ and $m_{I}^{*}=20.90$. The dotted curve shows the magnitude distribution of the 563 galaxies nearest the $z=0.5$ cluster red sequence in the BTC data, while the star symbols indicate the magnitude distribution of stars redder than the $z=0.5$ cluster red sequence.}
\label{nummag}
\end{figure}

Figure~\ref{nummag} shows the excess of red galaxies binned by $I$ magnitude as a solid curve connecting filled circles. This is determined by scaling the magnitude distribution of galaxies in the DLS image to have the same effective area as the BTC field, and subtracting it from the magnitude distribution of red galaxies in the BTC field. Significant excesses are observed for all bins in the range \mbox{$20<I_{tot}<23.5$}. The excess is particularly significant for \mbox{$20<I_{tot}<21$} where 364 red galaxies are found in the BTC field whereas only 209 would be expected from the DLS data, corresponding to an excess of 74\%, and a significance level of \mbox{4.3$\,\sigma$}, assuming the same effect on statistics of galaxy clustering as previously.

For comparison the magnitude distribution of stars redder than the $z=0.5$ cluster red sequence (i.e. having the same colours as the red galaxies) in the BTC field is indicated by the star symbols. The magnitude distribution of red stars appears flat, and is quite distinct from that of the red galaxy excess, again confirming that the red galaxy excess is not primarily due to stellar contamination. In particular, at bright magnitudes \mbox{($I<21$)} where the stellar classifier is virtually 100\% efficient, the two magnitude distributions are completely different.

To examine whether the excess could be due to a large photometric error producing a colour shift in either the BTC or DLS data, we introduce an artificial shift in the \mbox{$V-I$} colours of the BTC galaxies sufficiently large to remove the observed excess in the BTC field. To account for the whole excess in this manner requires a \mbox{$V-I$} colour shift of -0.15\,mag. Assuming zeropoint calibration uncertainties of 0.01\,mag for both $V$ and $I$ DLS images, we obtain an overall relative photometric calibration uncertainty between the BTC and DLS data of \mbox{$\Delta(V-I)=0.03$}. Hence, a calibration error as large as 0.15\,mag appears unlikely, but as a further check we consider the magnitude distribution of those galaxies which are initially classed as red galaxies, but which would be lost if there was a \mbox{$V-I$} colour shift of -0.15\,mag, i.e. those with
\begin{equation}
\begin{array}{c}
3.689 - 0.0652\times I < V-I <  3.839 - 0.0652\times I\;; \\ 
I_{total}<23.5.
\end{array}
\end{equation} 
The magnitude distribution of these galaxies is shown in Fig.~\ref{nummag} as a dotted curve, and is clearly substantially different to the magnitude distribution of the observed excess, the former being unable to reproduce the sharp increase at \mbox{$I\simeq20$} observed in the latter. The distinctness of both the colour and magnitude distributions of the observed red galaxy excess in comparison to those predicted for excesses produced by stellar contamination or photometric calibration errors, in particular the peak at \mbox{$V-I\simeq2.8$} and the sharp increase at \mbox{$I\simeq20$}, indicates that the red galaxy excess observed is not primarily due to either of these sources.

Given the relative ease of differentiating between real and spurious sources in the image, that although their removal is a subjective process, we estimate the uncertainty resulting from this process to be at the 1\% level, or $\sim25$. The major uncertainties in the final figure are the level of stellar contamination, which could contribute as many as $\sim250$ sources, and the uncertainty in the zero-point of the $V-I$ colour, which we take to be \mbox{$\Delta(V-I)=0.03$}.  resulting in an uncertainty in the red galaxy numbers of 117 (the number of red galaxies which would be lost if their \mbox{$V-I$} colour were reduced by 0.03\,mag.) The effect of uncertainty in the $I$-band magnitude is much less, even assuming a pessimistic level of \mbox{$\Delta I=0.02$} produces an uncertainty of only 18 red galaxies (the number of galaxies with \mbox{$23.48<I<23.50$}).

The galaxy luminosity function is a powerful tool in cosmology by allowing the bulk properties of galaxies to be measured. Numerous studies show that the luminosity function is well described by the Schechter (\cite{schechter}) function
\begin{equation}
\phi(L)dL = \phi^{*}\left(\frac{L}{L^{*}}\right)^{\alpha}e^{-(L/L^{*})}\,d\left(\frac{L}{L^{*}}\right),
\end{equation} 
which is written as a function of apparent magnitude as
\begin{equation}
\phi(m)dm = k\phi^{*}e^{[-k(\alpha+1)(m-m^{*})-e^{-k(m-m^{*})}\,]}\;dm,
\end{equation}
where  $k=0.4\ln 10$.

\begin{figure*}
\centerline{{\resizebox{17cm}{!}{\includegraphics{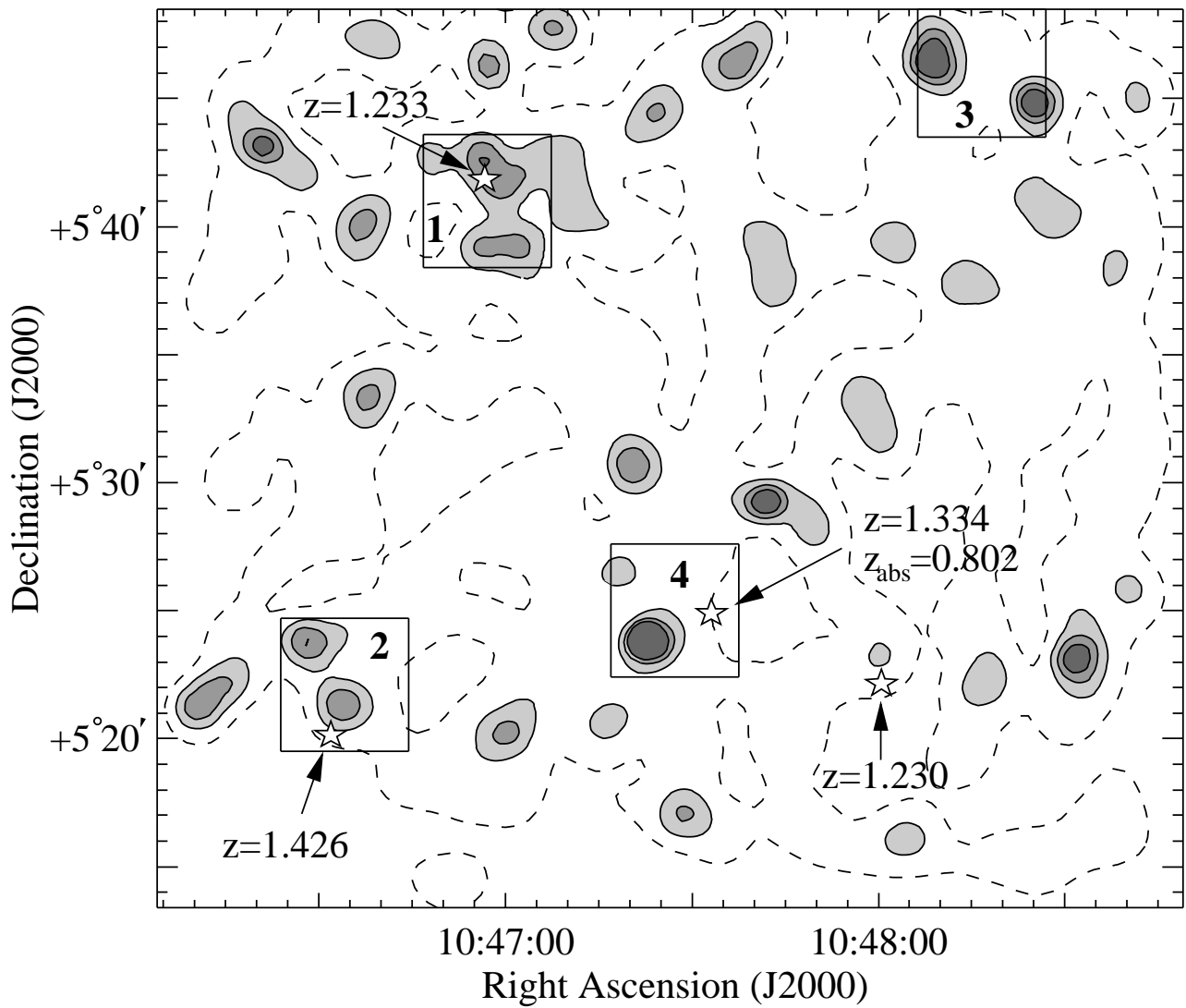}}}}
\caption{Contour plot of the estimated density distribution of {\em red galaxies} in the BTC field (corresponding to \mbox{$31\times27\sqmpc$} at \mbox{$z=1.2$}). The dashed contour corresponds to the density of 1.65 galaxies arcmin$^{-2}$, that observed for the DLS data. The shaded regions indicate the density peaks with solid contours corresponding to overdensities of 100\%, 200\% and 300\% in comparison to the DLS data. The four quasars at \mbox{$z\sim1.3$} are indicated by star symbols labelled by redshift. The NTT fields that targeted the four best cluster candidates at $z\sim1$ are indicated by the labelled boxes.}
\label{red_spatial}
\end{figure*}

The redshift evolution of the galaxy luminosity function has been examined for cluster populations out to \mbox{$z\sim1$} (e.g. de Propris \etal \cite{depropris}), finding that $L^{*}$ evolves in a manner consistent with the passive evolution of galaxies formed at early epochs, although there remains some debate over the effect of galaxy mergers on $\phi(L)$. Studies measuring the evolution of $m^{*}$ with redshift for galaxy clusters to \mbox{$z\sim1$} (e.g. Nelson \etal \cite{nelson}; de Propris \etal \cite{depropris}) indicate that $m^{*}(z)$ can be fit by a single function with intrinsic scatter between clusters at the same redshift of \mbox{$\sigma(m^{*})\sim0.2$}. This then affords the possibility of estimating the redshift of a cluster from its galaxy luminosity function, and Nelson \etal (\cite{nelson}) indicate that the rms error of redshift estimates from $m^{*}_{I}(z)$ for 44 clusters at \mbox{$0.3<z<0.9$} is \mbox{$\Delta(z)=0.06$}. By calculating the luminosity function of  \mbox{$\sim75\,000$} galaxies from the 2dF Galaxy Redshift Survey for different subsets defined by their spectral properties, Madgwick \etal (\cite{madgwick}) show that $\alpha$ varies significantly with spectral type. They find a systematic steepening of the faint-end slope, $\alpha$, from passive (\mbox{$\alpha=-0.54\pm0.02$}) to active star-forming (\mbox{$\alpha=-1.50\pm0.03$}) galaxies. 

We fit the Schechter function to the observed excess through a chi-squared minimization process, allowing $m_{I}^{*}$, $\phi^{*}$ and $\alpha$ to vary. The best-fitting function is shown as the dashed-curve in Fig.~\ref{nummag} and the optimal values for the free parameters found to be \mbox{$m_{I}^{*}=20.90\pm0.12$} and \mbox{$\alpha=-0.3\pm0.25$}. 

The observed shallow faint-end slope is comparable to that for local early-type galaxies (Madgwick \etal \cite{madgwick}), and hence confirms previous suggestions that the excess is made up of luminous early-type galaxies. It also appears to indicate that the excess is due to a structure at a single redshift, as a wide spread of redshifts would produce a steeper faint end curve as the luminosity function is convolved with the redshift distribution. 

In determining $m^{*}_{I}(z)$ Nelson \etal (\cite{nelson}) use a faint-end slope of \mbox{$\alpha=-1.25$}, and to compare our data with theirs, and to estimate the redshift of the excess galaxies we find that for a fixed \mbox{$\alpha=-1.25$} we obtain \mbox{$m_{I}^{*}=20.12\pm0.12$}, which results in a redshift estimate of \mbox{$z=0.83\pm0.08$}. It should be noted that the excess magnitude distribution we are measuring is colour selected and biased towards early-type galaxies, and this explains the shallow faint-end slope obtained, whereas Nelson \etal (\cite{nelson}) consider the {\em overall} galaxy excess as a function of magnitude. However as early-type galaxies dominate the bright-end of the galaxy luminosity function, this should not affect the redshift estimate significantly.

\begin{figure*}[t]
\centerline{{\resizebox{17cm}{!}{\includegraphics{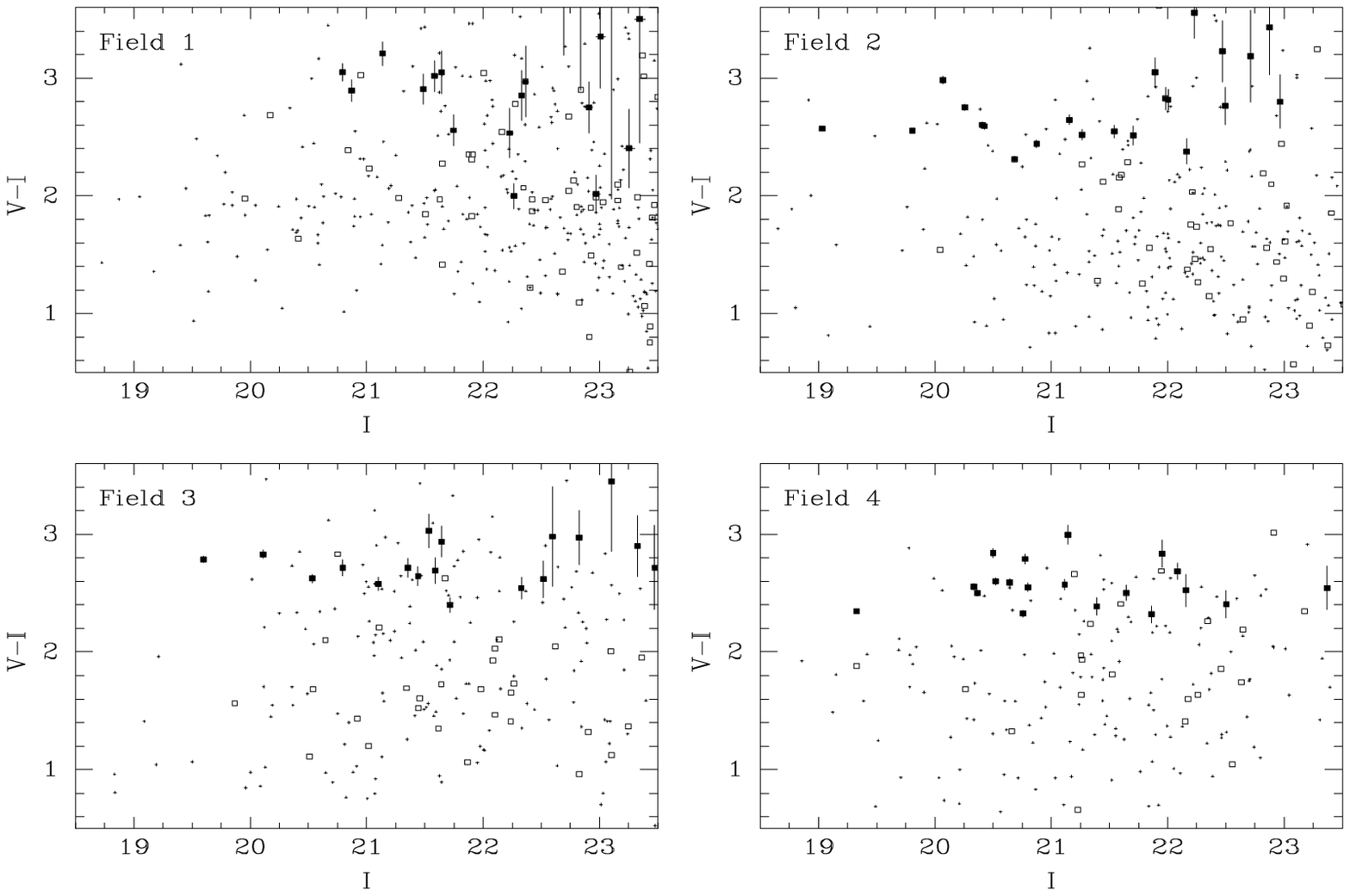}}}}
\caption{The $V-I$ against $I$ colour-magnitude diagrams for galaxies in the four NTT fields. Galaxies in the circular regions corresponding to the clusters identified in each field are indicated by squares, with those having (not having) the colours expected of early-type galaxies at the cluster redshift indicated by solid (open) symbols. For field 1 this corresponds to \mbox{$z\simeq1.2$}, and the galaxies are selected to have \mbox{$V-I>1.8$} and \mbox{$3.6<I-K_{s}<4.8$}, whereas for fields 2, 3 and 4 this corresponds to \mbox{$z\simeq0.8$} and the galaxies are selected to have \mbox{$V-I>2.3$} and \mbox{$3.00<I-K_{s}<3.75$}. The remaining galaxies in the fields are indicated by small crosses.}
\label{opticalCMs}
\end{figure*}

The fact that we are able to fit a Schechter function to the magnitude distribution of the excess, plus the observed peak in the colour distribution, indicates that this excess is predominately due to a superstructure containing many luminous galaxies at a single redshift of \mbox{$z\simeq0.8$}.  This corresponds to the redshift of one of the two structures identified in the quasar and Mg{\sc ii} absorber distributions, suggesting they are all manifestations of the same underlying large-scale structure.

\subsection{Spatial distribution of the red galaxies -- cluster candidates at $z\sim1$}

The overall excess of red galaxies across the BTC field is most likely due to the presence of one or more large-scale structures at \mbox{$z\sim1$}. As discussed previously, the most natural explanation is that either or both of the structures identified in the quasar and \mgii absorber distribution, the Clowes-Campusano LQG at \mbox{$z\simeq1.2$} and the structure at \mbox{$z\simeq0.8$}, bely the presence of an underlying large-scale structure in the form of galaxies. To examine any relation between the red galaxy excess and the Clowes-Campusano LQG, and to identify suitable cluster candidates for follow-up near-infrared studies, the density distribution of the red galaxies (i.e. those satisfying Eq. 1) across the BTC field is estimated through an adaptive kernel approach (Pisani \cite{pisani93}, \cite{pisani96}; Haines \cite{hainesphd}), as shown in Fig.~\ref{red_spatial}. The four quasars at \mbox{$z\sim1.3$} are indicated in the figure by star symbols labelled by redshift.

Four good \mbox{$z\sim1$} cluster candidates are identified across the BTC field as the most significant density enhancements in the red galaxy spatial distribution, and also having apparent red sequences at \mbox{$V-I\simeq2.8$}, thus minimising the possibility of projection effects. These were targetted for NIR imaging as shown in Fig.~\ref{red_spatial} by labelled boxes. Fig.~\ref{opticalCMs} shows the \mbox{$V-I$} against $I$ colour magnitude diagrams for galaxies in the NTT fields covering the four best \mbox{$z\sim1$} cluster candidates. Galaxies in the circular regions corresponding to the clusters identified in each field (see Fig.~\ref{spatial}) are indicated by squares, with those having (not having) the colours expected of early-type galaxies at the cluster redshift (based on the $VIK_{s}$ colours) indicated by solid (open) symbols. For field 1 this corresponds to \mbox{$z=1.2$}, and the galaxies are selected to have \mbox{$V-I>1.8$} and \mbox{$3.6<I-K_{s}<4.8$}, whereas for fields 2, 3 and 4 this corresponds to \mbox{$z=0.8$} and the galaxies are selected to have \mbox{$V-I>2.3$} and \mbox{$3.00<I-K_{s}<3.75$} (see Sect. 5.2). The remaining galaxies in the fields are indicated by small crosses. Each of the four NTT fields have excesses of 20--50 red galaxies in comparison to the DLS data, and between them account for 140 excess red galaxies, or 25\% of the total over the BTC field. Hence the clusters are not the sole reason for the excess, and the majority is due to a general large-scale excess. The only other cluster candidate ($10^\mathrm{h}48^\mathrm{m}31^\mathrm{s}$, $+05\degr 23\arcmin$) with a comparable excess of  ($\simeq20$) red galaxies has a red sequence at $V-I\sim2.4$ suggesting it is at a lower redshift.

\section{NEAR INFRARED IMAGING}

Photometric calibration of the NIR data was performed through the repeated observations of the faint near-infrared NICMOS standard stars of Persson \etal (\cite{persson}). In total 24 standards were observed through the 2 nights, matching the variation of airmass of the science exposures, resulting in zeropoints for each set of science data accurate to \mbox{$\sim0.01$\,mag} for both $J$ and $K_{s}$-bands. The $V$-, $I$-, and $J$-band data were registered with the $K_{s}$-band data, and the images convolved with a Gaussian kernel to match the PSF of the image with the worst seeing. Object detection was performed on the $K_{s}$-band image using {\sc SExtractor} (Bertin \& Arnouts \cite{bertin}), and colours determined using fixed apertures of 2.5\,arcsec diameter with {\sc SExtractor} in two-image mode. Photometric uncertainties were determined from the noise levels in the pre-smoothed images.

The coordinates and magnitude limits of each of the fields are presented in Table~\ref{photometry}. Field 1 corresponds to the galaxy cluster associated with the \mbox{$z=1.233$} LQG quasar, which had been previously identified from $K$ imaging (Haines \etal \cite{haines01a}). Fields 2 and 3 both contain double clusters identified as overdensities in the red galaxy spatial distribution in the BTC field. Field 4 contains a cluster previously identified from $K$ imaging to be at \mbox{$z\simeq0.8$} (Haines \etal \cite{haines01b}; Haines \cite{hainesphd}) as well as a \mbox{$z=1.334$} quasar from the LQG.

The ultimate aim of the near-infrared imaging is to identify galaxies with each cluster candidate irrespective of star-formation history to $M^{*}+2$ through photometric redshift estimates based on the $VIJK_{s}$ data. Here we only consider the early-type galaxies which allow the redshift and extent of the clusters to be determined through the red sequence method.

\subsection{Colour-Magnitude Relations}

\begin{figure}
\centerline{{\resizebox{\hsize}{!}{\includegraphics{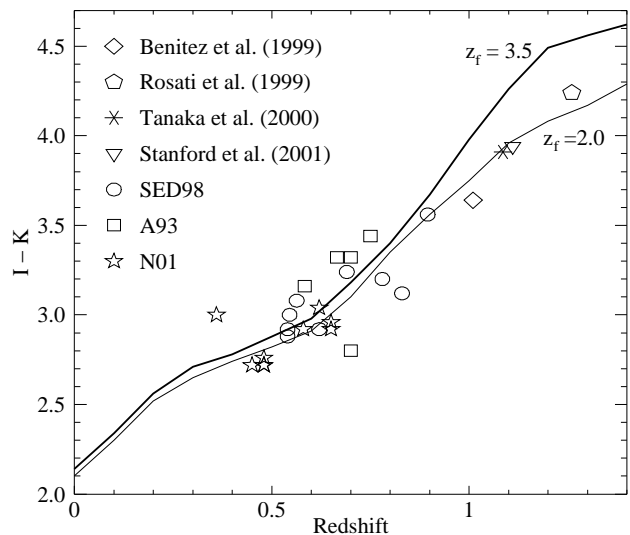}}}}
\caption{The $I-K_{s}$ colours of the cluster red sequence as a function of redshift. The solid curves predict the colour evolution of early-type galaxies as modelled by stellar populations formed in an instantaneous burst at $z_{f}=3.5$ (thick) and $z_{f}=2.0$ (thin). The points correspond to the colour of the red sequence for several known clusters at $z\simeq1$ as well as clusters from the surveys of SED98, A93 and N01.}
\label{I-K_cols}
\end{figure}

Figure~\ref{I-K_cols} shows the \mbox{$I-K_{s}$} colour of the cluster red sequence as a function of redshift. The solid curves show the predicted colour evolution of early-type galaxies as modelled by stellar populations formed in an instantaneous burst at \mbox{$z_{f}=3.5$} (thick) and \mbox{$z_{f}=2.0$} (thin). The points correspond to the colour of the red sequence for several known clusters at \mbox{$z\sim1$} (Ben\'{\i}tez \etal \cite{benitez}; Rosati \etal \cite{rosati}; Stanford \etal \cite{stanford02}; Tanaka \etal \cite{tanaka00}) and those from the surveys of SED98, A93 and N01. It is apparent that \mbox{$I-K_{s}$} photometry allows the redshift of clusters out to \mbox{$z\simeq1.3$} to be determined efficiently, with red sequences expected at \mbox{$I-K_{s}\simeq3.4$} and \mbox{$I-K_{s}\simeq4$.1--4.5} for clusters at \mbox{$z=0.8$} and \mbox{$z=1.2$} respectively.

\begin{table}
\begin{center}
\begin{tabular}{ccccccc}\hline\hline
\multicolumn{3}{c}{Field}	& \multicolumn{4}{c}{5$\sigma$ limit mag.arcsec$^{-2}$} \\
\#	& RA(J2000) 	& Dec(J2000) 	& $K_{s}$ & $J$	& $I$	& $V$\\ \hline
1 & $\!10^\mathrm{h}46^\mathrm{m}57\fs 1\!$ & $\!+05\degr41\arcmin00\arcsec\!$ & 21.1 & 23.4 & 25.7 & 26.4 \\
2 & $\!10^\mathrm{h}46^\mathrm{m}34\fs 2\!$ & $\!+05\degr22\arcmin09\arcsec\!$ & 21.1 & 23.0 & 26.2 & 26.8 \\
3 & $\!10^\mathrm{h}48^\mathrm{m}16\fs 6\!$ & $\!+05\degr46\arcmin07\arcsec\!$ & 21.1 & 23.2 & 26.3 & 26.6 \\
4 & $\!10^\mathrm{h}47^\mathrm{m}27\fs 3\!$ & $\!+05\degr24\arcmin59\arcsec\!$ & 20.3 & 22.6 & 25.9 & 26.8 \\
\hline
\end{tabular}
\end{center}
\caption{Coordinates and magnitude limits for the fields covered by near-infrared imaging.}
\label{photometry}
\end{table}

\begin{figure*}[t]
\centerline{{\resizebox{13cm}{10cm}{\includegraphics{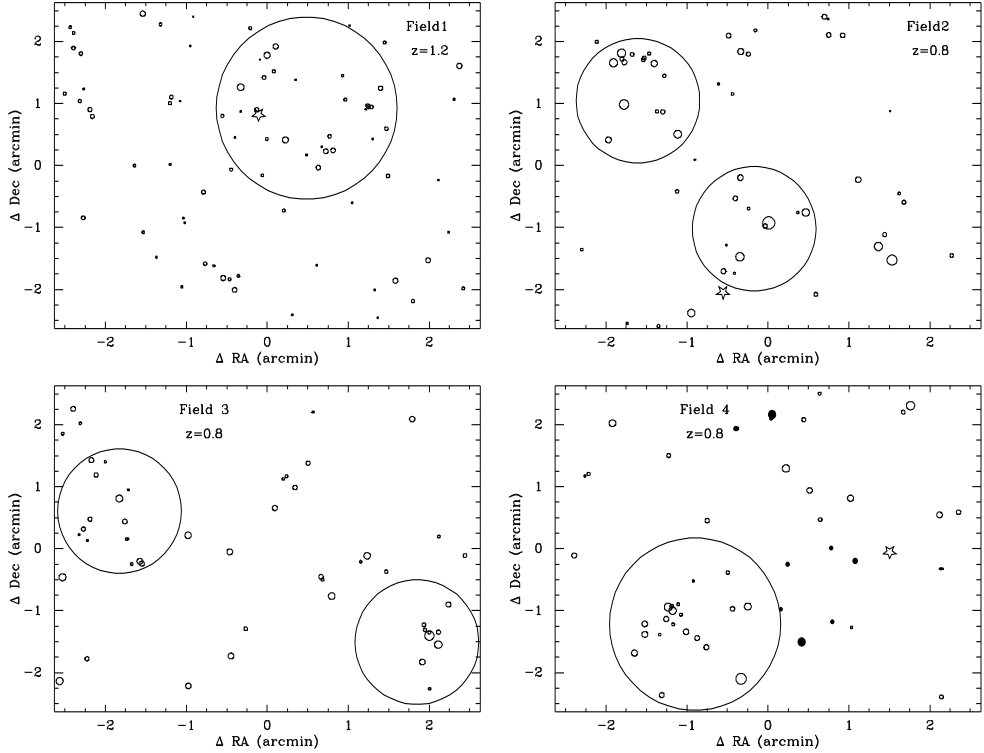}}}}
\caption{Spatial distribution of galaxies in the four NTT fields. Each open circle indicates the location of a galaxy with the colours expected of early-type galaxies at the redshift of the cluster identified in each field, whose size indicates the galaxy's $K_{s}$ magnitude, the area being proportional to the observed flux. For field 1 this corresponds to \mbox{$z\simeq1.2$}, and the galaxies are selected to have \mbox{$V-I>1.8$} and \mbox{$3.6<I-K_{s}<4.8$}, whereas for fields 2, 3 and 4 this corresponds to \mbox{$z\simeq0.8$} and the galaxies are selected to have \mbox{$V-I>2.3$} and \mbox{$3.00<I-K_{s}<3.75$}. In field 4 passively-evolving galaxies at \mbox{$z\simeq1.2$} are indicated by solid symbols. The nominal cluster regions used in Fig.~\ref{CMs} are shown by large circles. The location of each quasar is indicated by a star symbol.}
\label{spatial}
\end{figure*}

\begin{figure*}[t]
\centerline{{\resizebox{17cm}{!}{\includegraphics{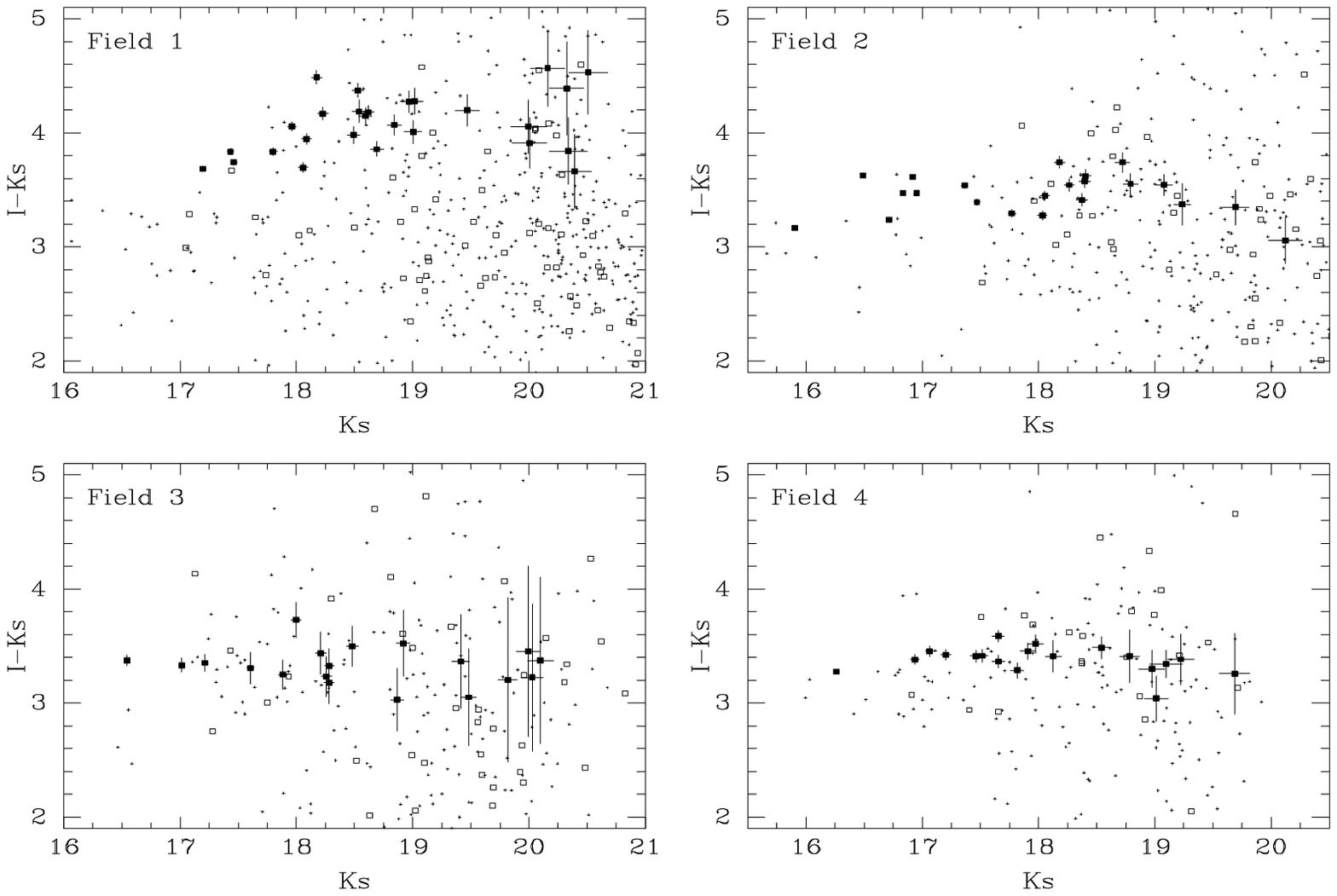}}}}
\caption{The $I-K_{s}$ against $K_{s}$ colour-magnitude diagrams for galaxies in the four NTT fields. Galaxies in the circular regions corresponding to the clusters identified in each field are indicated by squares, with those having (not having) the colours expected of early-type galaxies at the cluster redshift indicated by solid (open) symbols. For field 1 this corresponds to \mbox{$z\simeq1.2$}, and the galaxies are selected to have \mbox{$V-I>1.8$} and \mbox{$3.6<I-K_{s}<4.8$}, whereas for fields 2, 3 and 4 this corresponds to \mbox{$z\simeq0.8$} and the galaxies are selected to have \mbox{$V-I>2.3$} and \mbox{$3.00<I-K_{s}<3.75$}. The remaining galaxies in the fields are indicated by small crosses.}
\label{CMs}
\end{figure*}

We identify passively-evolving galaxies at \mbox{$z\simeq0.8$} as having \mbox{$V-I>2.3$} and \mbox{$3.0<I-K_{s}<3.75$}. These criteria were chosen partly {\em a posteriori} to cover the observed \mbox{$VIK_{s}$} colours of {\em red galaxies} in fields 2, 3 and 4, but a comparison with Figs.~\ref{modelcols} and~\ref{I-K_cols} shows that these criteria effectively select passive-evolving galaxies at \mbox{$0.8\pm0.2$}, the uncertainty corresponding to that expected for a redshift estimate for a galaxy at \mbox{$z\sim0.8$} from its $VIJK_{s}$ colour.

Passively-evolving galaxies at \mbox{$z\simeq1.2$} are identified in the NTT fields as those having \mbox{$V-I>1.8$} and \mbox{$3.6<I-K_{s}<4.8$}. The validity of these criteria can be confirmed, in a similar way as above, through comparison with the predicted colours of passively-evolving galaxies at \mbox{$1.2\pm0.2$} shown in Figs.~\ref{modelcols} and~\ref{I-K_cols}, but also with the results of the Gemini Deep Deep Survey (GDDS; Abraham \etal \cite{gdds}). GDDS aimed at obtaining a mass-selected sample of galaxies at \mbox{$1<z<2$}, by preferentially selecting those near the \mbox{$I-K_{s}$} versus $I$ colour-magnitude track mapped out by passively evolving galaxies in this redshift interval, and succeeded to secure high-quality spectra of quiescent galaxies at \mbox{$1<z<2$}. They identified twelve passively-evolving galaxies at \mbox{$1.2<z<1.4$}, all of them having \mbox{$V-I>1.8$} and \mbox{$3.5<I-K_{s}<4.8$}. In contrast all of their \mbox{$z<0.8$} galaxies, excepting one, have \mbox{$I-K_{s}<3.6$}. The characterisation of a bluer \mbox{$V-I$} selection criterion for \mbox{$z\simeq1.2$} galaxies with respect to those at \mbox{$z\simeq 0.8$} can be qualitatively understood from Fig.~\ref{modelcols}. In effect, the change on the galaxy \mbox{$V-I$} colours due to the presence of star-formation, even in a small amount, is much larger for the \mbox{$z\simeq1.2$} galaxies; this is apparent from the divergence at \mbox{$z\simeq1.2$} of the curves corresponding to the models describing passive and disk-dominated galaxies at \mbox{$z\simeq1.2$}. Such behaviour is confirmed by the GDDS, as five of the twelve galaxies at \mbox{$1.2<z<1.4$} spectroscopically-classified as passively-evolving have \mbox{$1.8<V-I<2.3$}. We use a slightly redder minimum in our \mbox{$I-K_{s}$} selection than the one resulting from the GDDS, in order to minimise the foreground contamination expected in our case from galaxies at \mbox{$z\simeq0.8$}.

Figure~\ref{spatial} shows the spatial distribution of those galaxies in the four NTT fields identified as early-type galaxies at the redshift of the cluster located in each field (\mbox{$z\simeq1.2$} for field 1; \mbox{$z\simeq0.8$} for fields 2, 3 and 4) from their photometry. Each galaxy is indicated by an open circle, whose size indicates the galaxy's $K_{s}$ magnitude, the area being proportional to the observed flux. The large circles indicate the nominal cluster regions used in Figs.~\ref{opticalCMs} and~\ref{CMs}. The location of each quasar is indicated by a star symbol. 

Figure~\ref{CMs} shows the $I-K_{s}$ against $K_{s}$ colour-magnitude diagrams for galaxies in the four NTT fields. To highlight the signal from the cluster red sequences in each of the fields, circular regions around each of the nominal cluster centres are defined as shown in Fig.~\ref{spatial}. For fields 1 and 4, one circular region of radius $1'$ is defined, and for fields 2 and 3 where there is an apparent bimodal structure, 2 circular regions of radius $(2^{-1/2})'$ are defined. Galaxies within these circular regions are indicated by squares, with those having (not having) the colours of early-type galaxies at the cluster redshift indicated by solid (open) symbols. The remaining galaxies within the NTT fields are indicated by small crosses. 

Field 1 contains the \mbox{$z\simeq1.2$} cluster previously identified by Haines \etal (\cite{haines01a}), and this is confirmed by the excess of galaxies with \mbox{$I-K_{s}\simeq4$} in the circular region around the nominal cluster centre. This is as expected given that the circular region corresponds approximately to the \mbox{$2.25\times2.25\sqarcmin$} UFTI field used to obtain the $K$ imaging described in Haines \etal (\cite{haines01a}). Although a tight red sequence is not apparent, 28 $K_{s}<20.5$ galaxies within 1\,arcmin of the cluster centre are classed as early-type galaxies at \mbox{$z\simeq1.2$} from their photometry, whereas only 2--4 would be expected. Figure~\ref{spatial} (top-left) shows that the clustering appears significantly extended beyond that identified by Haines \etal (\cite{haines01a}), with \mbox{$\sim80$} galaxies having colours consistent with passively-evolving galaxies at \mbox{$z\simeq1.2$} across the NTT field, forming an amorphous, clumpy structure \mbox{4--5$\mpc$} in extent.

In each of the three remaining fields, clear red sequences of \mbox{$\simeq20$} galaxies are apparent at \mbox{$I-K_{s}\simeq3.$4--3.5} in the colour-magnitude distributions of galaxies in the circular regions around the nominal cluster centres. Indeed the red sequences for the clusters in fields 3 and 4 appear almost identical, with that of field 2 marginally redder (\mbox{$\Delta I-K_{s}\sim0.1$}), and comparison with Fig.~\ref{I-K_cols} indicates that all three clusters are at \mbox{$z=0.8\pm0.1$}. The red sequence galaxies in fields 2 and 3 appear bimodally distributed, with two compact groups of galaxies located 3 and 5 arcmin apart respectively (Fig.~\ref{spatial}). In contrast the distribution of red sequence galaxies in field 4 appears unimodal, forming a single compact cluster with no obvious large-scale filamentary structure elsewhere in the field.

In field 1, as well as the clustering at \mbox{$z\simeq1.2$} manifested by the galaxies with \mbox{$I-K_{s}\simeq4.1$} (Fig.~\ref{CMs}; top-left), there appears a second distinct red sequence at \mbox{$I-K_{s}\simeq3.4$}, identical in colour to those apparent in the remaining three fields, placing them at the same redshift. In contrast to fields 2, 3 and 4, the galaxies with \mbox{$I-K_{s}\simeq3.4$} in field 1 appear evenly distributed across the whole NTT field.

Although there is no further evidence of rich clusters at \mbox{$z\simeq1.2$} in the NTT data, in each of fields 2, 3 and 4, there are \mbox{$\sim10$} $K_{s}\lsim 19$ galaxies with the $VIJK_{s}$ colours consistent with L$^{*}$-class galaxies at $z\simeq1.2$ with predominately old stellar populations. These do not appear clustered, except for a group of \mbox{$\sim6$} galaxies with \mbox{$I-K_{s}\simeq4.1$} located near the \mbox{$z=1.334$} LQG quasar in field 4 (indicated by solid symbols in Fig.~\ref{spatial}), suggesting that the latter is associated with this group.

\subsection{Star-Galaxy Separation}

The addition of near-infrared imaging allows stars and galaxies to be distinguished on the basis of colour as well as morphology, with stars constrained to narrow loci in the \mbox{$J-K_{s}$} or \mbox{$I-K_{s}$} against \mbox{$V-I$} colour-colour diagrams, appearing much bluer in the near-infrared than galaxies of comparable optical colours. It is thus possible to identify stars on the basis of their near-infrared colours, and to examine the efficiency of the morphological classification and estimating the level of stellar contamination in the red galaxy subset. We find that 10 of the 122 sources in the three deepest NTT fields which are classified as stars on the basis of morphology have the near-infrared colours of galaxies; these are mostly faint \mbox{($I\gsim 23$)} blue galaxies. Of the 227 red galaxies in NTT fields 1--3 (field 4 is not used due to its insufficient depth), 17 are found to have the near-infrared colours of stars. By assuming that this density of stellar contamination is maintained over the whole BTC field, we would expect a total stellar contamination level of \mbox{$250\pm65$} or 10\%, where the quoted uncertainty assumes Poisson noise for the 17 red galaxies identified as stars in the three NTT fields. As the image quality of the BTC and DLS images are similar in terms of depth and seeing levels, and lie at similar galactic latitudes (+53\degr\, for BTC; -50\degr\, for DLS), we would expect a similar level of stellar contamination in the DLS data. Therefore, although the number appears significant in comparison to the total excess, the uncertainty in the relative levels of stellar contamination between the two sets of data is just \mbox{$\sigma\approx65\times\sqrt{2}\approx90$}, and so it is unlikely that more than 20\% of the 563 galaxy excess (see section 4.2) is due to stellar contamination, particularly as the magnitude and colours distributions of the red galaxy excess and stellar populations are so different.

\section{DISCUSSION}

The realization that coherent massive structures \mbox{$\gsim 100\mpc$} in size exist in any sufficiently large volume in the universe up to at least to \mbox{$z\sim1$} would provide a new major constraint for current and future theoretical models of the universe and its evolution. It is well known and widely accepted that structures \mbox{50--100$\mpc$} in diameter exist in the local universe, however at higher redshifts it becomes much more difficult to detect similar structures and eventually we may be biased to detect mostly the largest ones. Very recently, two \mbox{100$\mpc$}-scale structures at \mbox{$z\sim0.3$} were detected in the three-dimensional distribution of radio galaxies in a \mbox{25\,deg$^2$} area (Brand \etal \cite{brand}). In another \mbox{25\,deg$^2$} region of the sky, previous investigations have identified two \mbox{100$\mpc$}-scale structures at \mbox{$z\simeq0.8$} and \mbox{$z\simeq1.2$} (Clowes \& Campusano \cite{clowes91}, \cite{clowes94}; Williger \etal \cite{williger}), using counts of quasars and \mgii absorbers. It is not completely clear whether the volumes surveyed in these surveys are typical, or rather special, containing one or more rare fluctuations in the density distribution of the objects studied. In the case studied first by Clowes \& Campusano(\cite{clowes91}), although there are reasons to believe that an excess in the number of quasars is marking a mass overdensity, the fact that it is also marked by an excess of \mgii absorbers is more directly indicative of an accompanying galaxy overdensity. In order to further characterize the distribution of the galaxy overdensity implied by \mbox{100$\mpc$}-scale structures in the Clowes \& Campusano(\cite{clowes91}) area, we have undertaken an ultra-deep optical imaging survey in a subarea of \mbox{$40\times35\sqarcmin$} -- denominated the BTC field --, to search for evidence of galaxy overdensities at $z\simeq0.8$ and $z\simeq1.2$, either in the form of clusters, structures or a uniform distribution of galaxies at each redshift. Our analysis of the BTC data provided suggestive evidence for large-scale structure in this area of the sky, and allowed the identification of four cluster candidates all at \mbox{$z\sim1$}. Next, we obtained targeted NIR imaging around these cluster candidates to ascertain the actual nature of the clusters and the large-scale structures, and to evaluate if they can be related with the quasar and \mgii absorber superstructures. Using both the optical and NIR data, we obtain results showing that, at least in the probed volumes, the two $z\sim1$ $100\mpc$-scale structures do indeed present an enhanced density of galaxies. As the BTC field is a representative subarea of the region where the superstructures lie, we consider our results to be evidence for the presence of sheet-like galaxy structures at \mbox{$z\sim1$} and an indication that they may be ``Great Wall''-like structures at this early epoch of the universe. Although we describe the two $z\sim1$ structures as ``sheets'', given the sparse spatial sampling of the superstructures offered by the quasars and \mgii absorbers, and the lack of redshift information for the galaxies, we cannot accurately measure the topology of the superstructures. However, their extent over the plane of the sky indicates that they are at least two-dimensional (as opposed to filamentary) structures.

\subsection{Selection of early-type galaxies at 0.5$\lsim$z$\lsim$1.3}

To search for a possible large-scale structure of galaxies at \mbox{$z\sim1$} we have applied a variant of the cluster red sequence algorithm of Gladders \& Yee (\cite{gladders}) to ultra-deep $V$ and $I$ imaging data of the BTC field. Instead of considering narrow redshift-slices over the redshift range of interest, a single high-redshift slice is used, containing all \mbox{$I<23.5$} galaxies redder in \mbox{$V-I$} than the cluster red sequence at \mbox{$z=0.5$}. Simple galaxy evolution models predict that this selection of {\em red galaxies} effectively identifies a population of passively-evolving early-type galaxies at \mbox{$0.5\lsim z\lsim 1.3$}, whose strong 4000\AA\, breaks are responsible for their red \mbox{$V-I$} colour. The validity of this approach is demonstrated by application of the colour-magnitude criteria to galaxies in the EIS HDF-South field for which deep $UBVRI$ (and for 40\% of the field $JHK_{s}$) photometry exists. Of the 34 red galaxies identified: 32 were constrained at the \mbox{$1\sigma$} level to be at \mbox{$z>0.6$} from their photometric redshifts, the remaining two constrained to \mbox{$z>0.45$}; and all were best-fit by early-type galaxy models.

\subsection{A new large-scale structure of galaxies at z$\,\simeq\,$0.8}

Any large-scale galaxy structure at \mbox{$z\sim1$} underlying either of the quasar/\mgii superstructures at \mbox{$z\simeq0.8$} or \mbox{$z\simeq1.2$} should manifest itself as an excess of red galaxies across a sub-area under study such as the whole BTC field. In effect, through comparison with a suitable control field (F1p22 from the Deep Lens Survey of Wittman \etal \cite{wittman}) an excess of 563 red galaxies across the BTC field is identified. The magnitude and colour distributions of the excess red galaxies show unambiguously that the excess is largely due to a population of early-type galaxies at \mbox{$z\simeq0.8$}, and cannot be explained as either a calibration error or due to stellar contamination. The colour distribution of this excess shows a coherent peak at \mbox{$2.7<V-I<3.1$}, the signature predicted for a population of early-type galaxies at \mbox{$0.5\lsim z\lsim 1.3$}, and the magnitude distribution is well described by a Schechter function with a shallow-faint end slope (\mbox{$\alpha=-0.3\pm0.25$}) indicative of early-type galaxies (Madgwick \etal \cite{madgwick}). A comparison of $m^{*}_{I}$ with those of cluster populations out to \mbox{$z\simeq1$} (e.g. Nelson \etal \cite{nelson}) places the excess at \mbox{$z=0.83\pm0.08$}. 

The presence and redshift of this \mbox{$z\simeq0.8$} structure is confirmed by the targeted near-infrared imaging of the four best galaxy cluster candidates that had been identified in the BTC field, resulting in the identification of three clusters at $z=0.8\pm0.1$, each having clear red sequences of $\sim20$ galaxies with \mbox{$I-K_{s}\simeq3$.4--3.5}. Independent of the overall red galaxy excess, these clusters form a superstructure at \mbox{$z\simeq0.8$} that is \mbox{$\sim20\mpc$} across, which extends fully across the BTC field. 

There is no obvious apparent filamentary structure connecting the $z\simeq0.8$ clusters in the spatial distribution of red galaxies (Fig.~\ref{spatial}). However as the clusters make up only 20\% of the overall red galaxy excess, the most important contributor to the remainder is likely to be due to a connecting filamentary or sheet-like large-scale structure. Such a filamentary structure is difficult to detect from the red galaxy distribution, as it is likely to be diffuse, and the fraction of early-type galaxies will be lower than in the clusters, reducing the signal still further. Only further near-infrared imaging or multi-object spectroscopy over a wide area is likely to detect such filamentary structures. There is however evidence of such a diffuse, filamentary or sheet-like structure in field 1, where despite there being no clusters at $z\simeq0.8$, a red sequence at $I-K\simeq3.4$ is apparent, made up of galaxies evenly distributed across the NTT field. This suggests that the overall excess of red galaxies is due to two main components, on the one hand, the three $z\simeq0.8$ clusters, and on the other hand, a diffuse sheet-like structure extending across the whole BTC field.

\subsection{A new large-scale structure of galaxies at z$\,\simeq\,$1.2}

Previous $K$-band imaging of a \mbox{$2.25\times2.25\sqarcmin$} field centred on the \mbox{$z=1.233$} LQG quasar indicated the presence of associated clustering in the form of red sequences of 15--18 galaxies with the colours expected for a population of passively-evolving massive ellipticals at the quasar redshift (\mbox{$I-K\simeq4.3$}, \mbox{$V-K\simeq6.9$}). This field makes up part of NTT field 1 and as such has been reobserved with higher quality NIR data, with a field of view five times larger than before, having a magnitude greater depth in $K$, along with new $J$-band imaging. The previously observed red sequence is found again, with 28 $K_{s}<20.5$ galaxies within 1\,arcmin of the cluster centre, although now at \mbox{$I-K_{s}\simeq4.1$}. This appears due to a systematic offset between the two data sets, and given the extensive and well-behaved photometric calibrations of the NTT data, we consider these to be correct. It had been suggested in Haines \etal (\cite{haines01a}) that the clustering extended beyond the $K$ image, by consideration of the spatial distribution of optically red galaxies (Fig. 7 of Haines \etal \cite{haines01a}), with a filament predicted to extend 3\,arcmin to the east of the quasar, and a compact group 3\,arcmin to the south. These are covered by the larger NTT field, and are apparent in Fig.~\ref{spatial}(top-left), along with another group 2.5\,arcmin to the west of the quasar. The overall structure extends across the NTT field, corresponding to 3--4$\mpc$ at \mbox{$z\simeq1.2$}, and appears clumpy and filamentary. This appears to be a very massive system, considering its redshift, with $\sim80$ galaxies (40 with \mbox{$K_{s}<19$}) having colours consistent with being passively-evolving massive ellipticals at \mbox{$z\simeq1.2$}. There are also comparable numbers of galaxies having colours consistent with being star-forming galaxies at the same redshift (\mbox{$I-K_{s}>3.6$}, \mbox{$V-I<1.8$}) as seen also for other \mbox{$z\gsim1$} clusters (Tanaka \etal \cite{tanaka00}) and NIR surveys (McCarthy \etal \cite{mccarthy}).

Although there is no further evidence of rich clusters at \mbox{$z\simeq1.2$} in the BTC field, there is a group of \mbox{$\sim6$} galaxies with \mbox{$I-K_{s}\simeq4.1$} located near the \mbox{$z=1.334$} LQG quasar in field 4, suggesting that the latter is associated with this group. Also \mbox{$\sim10$} galaxies with \mbox{$I-K_{s}\simeq4.1$} are apparent in each of the NTT fields 2, 3 and 4, which could indicate a diffuse structure at \mbox{$z\simeq1.2$} across the BTC field. 

It is difficult to ascertain the full extent of the \mbox{$z\simeq1.2$} large-scale galaxy structure in the regions where only optical data exists, and to compare the structure with that of the \mbox{$z\simeq0.8$} structure. There are a number of effects which combine to significantly reduce the efficiency of the {\em red galaxy} selection criteria in detecting structures at \mbox{$z=1.2$} as opposed to \mbox{$z=0.8$}. Firstly the galaxies are 1--1.5 mag fainter due to the greater distance and the effects of k-corrections. Secondly the \mbox{$V-I$} colour samples shorter rest-frame wavelengths (from 3000--4500\AA\, at \mbox{$z=0.8$} to 2500--3600\AA\, at \mbox{$z=1.2$}), so the same amount of recent star-formation has approximately double the effect on the \mbox{$V-I$} colour of a galaxy at \mbox{$z=1.2$} as it would in the same galaxy at \mbox{$z=0.8$}. The galaxies themselves are younger (by \mbox{$1.2\,h^{-1}\,$Gyr}) and so are intrinsically bluer. Finally as the clusters themselves are younger and less dynamically evolved, their inhibitory effect on star-formation in their member galaxies is reduced, as is evident from the observations of the Butcher-Oemler (\cite{butcher}) effect. To date the redshift limit to which cluster red sequences have been found is $z\sim1.3$.

Many of these effects can be partially counteracted by the addition of NIR photometry, which is less affected by star-formation, allowing a greater fraction of the early-type galaxies to be detected. Of the \mbox{$\sim30$} galaxies in fields 2, 3 and 4 with \mbox{$I-K_{s}\sim4.1$} (and hence candidate early-type galaxies at \mbox{$z\simeq1.2$}), only \mbox{$\sim50$\%} would be classed as {\em red galaxies}, either appearing too blue in \mbox{$V-I$} (either due to some star-formation or photometric uncertainties as the limiting magnitude in $V$ is reached), or have \mbox{$I>23.5$} and are thus too faint. Hence approximately half the galaxies likely to be early-type galaxies at \mbox{$z\simeq1.2$} are missed by the optical selection criteria, and it is only the very rare  high-density peaks corresponding to rich clusters that become apparent at \mbox{$z\simeq1.2$} in the {\em red galaxy} spatial distribution. In contrast it is only in the regions with the NIR data that the sheet-like large-scale structure associated with the Clowes-Campusano LQG becomes apparent. This loss of efficiency in identifying early-type galaxies at \mbox{$z>1$} from optical data alone is apparent from the results of the Las Campanas Infrared Survey (McCarthy \etal \cite{mccarthy}) where galaxies identified as being the progenitors of early-type galaxies at \mbox{$1\lsim z\lsim 2$} from having \mbox{$I-H>3$} span a range in \mbox{$V-I$} colour of more than 3 mag, with around half of them having $V-I<2$.

\subsection{Large-scale structures at z$\,\lsim\,$0.3}

To date the distribution of galaxies on 100$\mpc$-scales has been followed out to \mbox{$z\sim 0.3$} (e.g. Geller \& Huchra \cite{geller}; Doroshkevich \etal \cite{d00}), which is also the upper limit of the major galaxy redshift surveys such as the 2dF\,GRS survey (Colless \etal \cite{colless}) and the  Sloan Digital Sky Survey (SDSS; York \etal \cite{york}). In consequence studies into the relation between quasars and the large-scale galaxy distribution have been limited to \mbox{$z<0.3$}. S\"{o}chting \etal (\cite{sochting}, \cite{sochtingphd}), although based on 2-colour photographic sky survey data, have presented not only evidence of large-scale structure at \mbox{$z\simeq0.3$} delineated by galaxy clusters but also that quasars occur preferentially close to these structures. Two \mbox{$100\mpc$} structures of radio galaxies have been recently discovered at \mbox{$z\sim0.3$} in a \mbox{25\,deg$^2$} area of the sky (Brand \etal \cite{brand}), which are estimated to be highly improbable under the quasi-linear structure formation theory and thus need to invoke an increased bias on the large scales.

\subsection{Large Quasar Groups as tracers of large-scale structure at high-redshifts}

The detection here of large-scale structures of galaxies that delineate part of known quasar/\mgii superstructures at \mbox{$z\simeq1$}, is similar to that found by Tanaka \etal (\cite{tanaka00}, \cite{tanaka01}) for the Crampton \etal (\cite{crampton89}) group of 23 quasars at \mbox{$z\simeq1.1$}. In a study comparable to that of Haines \etal (\cite{haines01a}), Tanaka \etal (\cite{tanaka00}) obtained deep $R$-, $I$- and $K$-band imaging of a \mbox{$5\times3.2\sqarcmin$} region centred on the \mbox{$z=1.086$} radio-loud quasar 1335.8+2834 from the Crampton \etal LQG. A rich cluster of galaxies with the colours predicted of passively-evolving galaxies at \mbox{$z\simeq1.1$} (\mbox{$I-K\simeq4.0$}, \mbox{$R-K\simeq5.3$}) is identified, and, as is found by Haines \etal (\cite{haines01a}), the quasar is located on the cluster periphery in a star-forming region, indicated by a band of emission-line galaxies (Hutchings \etal \cite{hutchings}), and also there is evidence of further filamentary structures forming a large-scale structure \mbox{4--5$\mpc$} in extent. 

In a wide-area \mbox{($48\times9\sqarcmin$)} imaging survey in both $R$- and $I$-bands toward a region containing five quasars from the Crampton \etal (\cite{crampton89}) LQG, Tanaka \etal (\cite{tanaka01}) detect significant clustering of faint, red galaxies with \mbox{$21<I<23.5$} and \mbox{$1.2<R-I<1.6$}, i.e. galaxies with the colours and magnitudes expected of early-type galaxies at \mbox{$z\simeq1.1$}. These galaxies are concentrated in 4--5 clusters forming a linear structure of extent \mbox{$\sim10\mpc$} that is traced by members of this large group of quasars, although only the one radio-loud quasar of Tanaka \etal (\cite{tanaka00}) appears directly associated with any of the rich clusters.
Conversely, in an optical spectroscopic follow-up of six X-ray sources detected in a \mbox{$20\times20\sqarcmin$} field, known to contain two rich galaxy clusters at \mbox{$z\sim1.3$} (RXJ0848.6+4453 at \mbox{$z=1.273$} and RXJ0848.9+4452 at \mbox{$z=1.261$}; Stanford \etal \cite{stanford97}; Rosati \etal \cite{rosati}), three of the sources were identified as quasars at \mbox{$z=1.260$}, \mbox{$z=1.286$} and \mbox{$z=1.260$} (Ohta \etal \cite{ohta}). 

There are two main factors which should affect how quasars trace mass: the requirement of some disturbance to the host galaxy to push gas onto the central nucleus and activate the quasar, either due to a merging event, which is likely to be the dominant mechanism at \mbox{$z\simeq1$}, or during the formation of the galaxy (Haehnelt \& Rees \cite{haehnelt}); and the requirement of a massive black-hole to fuel the quasar, which given the strong correlation between the masses of the black hole and the bulge of the host galaxy (Merritt \& Ferrarese \cite{merritt}), implies a massive host galaxy.
 
The effect of quasars being located in merging/forming galaxies on how they trace mass is likely to be complex and redshift dependent. However on large-scales the effect should be small, and the most important consequence should be the avoidance by quasars of the high-density cluster centres at least to \mbox{$z\sim1$}. This is understandable in the framework of both galaxy merger and galaxy formation quasar triggering mechanisms: the encounter velocities of galaxies in the centre of clusters are much greater than the internal velocity distributions, and so galaxy mergers become much less efficient at triggering nuclear activity (Aarseth \& Fall \cite{aarseth}); and the cluster cores are filled with shock-heated virialised gas that does not easily cool and collapse (Blanton \etal \cite{blanton99}), inhibiting both the formation of stars and galaxies (Blanton \etal \cite{blanton00}), and hence inhibiting quasar formation. This effect has been observed in the form of the preferential location of quasars on the peripheries of clusters (S\'{a}nchez \& Gonz\'{a}lez-Serrano \cite{sanchez99}, \cite{sanchez02}; Haines \etal \cite{haines01a}; S\"{o}chting \etal \cite{sochting}; Tanaka \etal \cite{tanaka00}).

The observation that luminous (\mbox{$M_{V}<-23.5$}) quasars (at least to \mbox{$z\sim1$}) are located in massive ellipticals (e.g. McClure \etal \cite{mclure99}; Kukula \etal \cite{kukula}), and the requirement of a massive host galaxy, indicates that on large-scales at least (\mbox{$\gsim5\mpc$}) quasars should trace mass in the same way as their host galaxies, a prediction that appears confirmed by comparison of the quasar and galaxy power-spectra (Hoyle \etal \cite{hoyle}). We should thus expect quasar superstructures to trace the same mass overdensities as the large-scale structures mapped by galaxies, as is suggested by comparison of the quasar and galaxy spatial distributions in the local universe (Longo \cite{longo}). 

The results presented here and elsewhere (Tanaka \etal \cite{tanaka01}; Ohta \etal \cite{ohta}) indicate that super-large scale structures (\mbox{$\sim100 \mpc$}) in the form of galaxies, comparable to the ``Great Wall'' in the local universe, are apparent at \mbox{$z\sim1$}, and that they can be discovered and traced well by quasar superstructures. This suggests that large-scale quasar surveys, such as the 2dF QSO survey, could be used to map the evolution of the super-large scale structure to redshifts well beyond the scope of current galaxy redshift surveys, reaching $z\simeq1$ or beyond. Statistically significant structures on scales \mbox{$>100\mpc$} have been identified in the 2dF QSO survey (Miller \etal \cite{miller04}), appearing rarely, but throughout the survey, and over the whole redshift range covered \mbox{($0.5<z<2.2$)}.

\subsection{Evolution of Large-scale Structures}

The finding of large-scale structures at \mbox{$z\sim1$} can be understood in the context of the Zel'dovich non-linear theory, as the perturbations in the initial density field collapse first to form pancakes -- the sheet-like structures of the LSS -- before then collapsing on the remaining two axes to form filaments and clusters. As clusters and filaments are apparent at \mbox{$z\sim1$}, the sheet-like structures which form before these, must also be substantially in place. This is confirmed by simulations, in which Doroshkevich \etal (\cite{d99}) find that for the present day wall-like structures where the overdensities of matter are a factor 5--10 over the mean, half the mass was already in structures with the same level of overdensities at \mbox{$z\sim1$}. This also can  be considered through the simple argument that the characteristic distance scales of large-scale structures (\mbox{50--10$0\mpc$}) are much greater than the distances galaxies are likely to move over the time-scales involved, (a galaxy with a peculiar velocity of \mbox{$100\,h^{-1}\,\mathrm{km\,s}^{-1}$} moves \mbox{$1\mpc$} over 10\,Gyr). Hence, those galaxies that make up the large-scale structures apparent today and which were formed at \mbox{$z>1$}, will still constitute the same large-scale structures at \mbox{$z\sim1$}. 

The observation of two sheet-like large-scale structures at \mbox{$z\sim1$} in the same field, in conjunction with the results of Tanaka \etal (\cite{tanaka01}) and Ohta \etal (\cite{ohta}), implies that such structures although rare, are not exceptional. The observed early formation and relatively slow evolution of the abundances of massive structures, clusters in particular, is one of the fundamental reasons for favouring the current $\Lambda$CDM models (e.g. Bahcall \& Bode \cite{bahcall02}). The Hubble Volume numerical simulations of Evrard \etal (\cite{evrard}) find a large cluster at \mbox{$z=1.04$} in a $\Lambda$CDM model which has a mass twice that of Coma, making it the largest cluster in their positive octant (PO) survey which covered an eighth of the sky to \mbox{$z=1.46$}. In numerical simulations Doroshkevich \etal (\cite{d99}) find that sheet-like large-scale structures with typical maximal extensions of $\sim3$0--5$0\mpc$ incorporate $\sim40$\% of matter at $z=0$, of which $\sim60$\% is in place by $z\sim1$.

\section{CONCLUSIONS}

The results presented here support the presence of two sheet-like large-scale galaxy structures extending across the \mbox{20--3$0\mpc$} scales explored by our optical data: one ``sheet'' at \mbox{$z\simeq0.8$} with three associated clusters; and a second ``sheet'' at \mbox{$z\simeq1.2$} with one embedded cluster. Therefore, our results confirm that, at least in the volumes probed, the two \mbox{$z\sim1$} structures mark out volumes with an enhanced density of galaxies. As the studied subarea is quite representative of the whole region where the superstructures lie, we consider these results to be evidence for the presence of large sheets of galaxies at \mbox{$z\sim1$}, and a clear indication that such sheets may extend over \mbox{$100\mpc$}-scales. Secondly, these results show that large quasar groups (LQGs) are reliable indicators of galaxy excesses at \mbox{$\sim$100\,Mpc}-scales. In particular, the Clowes \& Campusano (\cite{clowes91}) LQG, the largest structure at high redshifts yet found, is then indicative of a huge structure of galaxies (and thus of coherent mass) over a volume of \mbox{$\sim100\times200\times200\,h^{-3}\mathrm{Mpc}^{3}$}.
Although the observational determination of the spatial frequency of such fluctuations in the distribution of galaxies at \mbox{$z\sim1$} will require future generations of galaxy surveys, the Clowes-Campusano LQG provides an extraordinary laboratory for the study of the relation of structures at different scales; the interplay between radio-galaxies, quasars and gas; and the evaluation of the bias factor for different constituents and environments inside the superstructure. 
Thirdly, the handful of LQGs that have been discovered so far at \mbox{$0.2<z<2.0$} (see Graham \etal \cite{graham95}), suggest that superstructures of this size are indeed rare, corresponding to unusually high peaks in the primordial density field. Brand \etal's (\cite{brand}) discovery of two $100\mpc$-scale structures of radio galaxies at \mbox{$z\sim0.3$} may also correspond to related evolution of similarly rare peaks.

The detections of two $100\mpc$-scale structures in surveyed volumes such as that of Clowes \& Campusano and Brand \etal studies, was shown by the latter to be incompatible with DM models predictions if these volumes are typical. Brand \etal explained the observations as a consequence of an increased ``bias factor'' and concluded that structures \mbox{$100\mpc$} in size at \mbox{$z\sim0.3$} do not present a challenge to the standard inflationary $\Lambda$CDM model. 
Using a semi-analytic model which includes both galaxy and quasar formation, in which the quasar activity is fuelled by major galaxy mergers, Enoki \etal (\cite{enoki}) find that quasars at \mbox{$z\sim1.2$} are biased with respect to the underlying mass distribution, with \mbox{$b\sim2$}, and are found in a variety of environments from small galaxy groups to rich clusters, as is observed (e.g. Wold \etal \cite{wold00}, \cite{wold01}). Equally, the massive galaxies which we are observing here, and which host the quasars in the LQGs, should have been twice as ``biased'' at \mbox{$z=1$} as they are at \mbox{$z=0$}, having \mbox{$b\sim2$} (Blanton \etal \cite{blanton00}). These bias factors allow the finding of such superstructures of quasars and galaxies to be reconciled with $\Lambda$CDM models.
Further theoretical work on this topic, together with specific predictions for the size distribution of the super-large structures as a function of redshift, is highly desirable. By assuming a quasar ``bias'' redshift dependence similar to that described above Miller \etal (\cite{miller04}) find that the significant \mbox{$>100\mpc$-scale} fluctuations in the quasar distribution observed in the 2dF QSO survey are in good agreement with predictions of a standard $\Lambda$CDM model.

As the ability of infra-red cameras to cover large areas of sky (tens of deg$^2$) is expected to improve greatly over the coming years (e.g. UKIRT-WFCAM, VISTA), so will the possibility of doing multi-colour studies of the evolution and the {\em bias function} of super-large scale structure to \mbox{$z\sim1$} and beyond. Also the newest generation of wide field spectrographs having large multiplexing (100--4000), either built (ESO-VIMOS and NIRMOS, and Magellan-IMACS and Keck-DEIMOS) or planned (Gemini-Optical MOS), allow the planning of redshift surveys for the study of specific populations of galaxies belonging to the largest structures yet known.

\begin{acknowledgements}
We thank the referee for their comments which have helped us to significantly improve the manuscript. CPH gratefully acknowledges the hospitality of the Universidad de Chile during his stay in Chile, where most of the data reductions were performed, and PPARC for travel support to La Silla. CPH acknowledges the financial support provided through the European Community's Human Potential Program under contract HPRN-CT-2002-00316, SISCO.
\end{acknowledgements}


\begin{thebibliography}{}
\bibitem[1980]{aarseth}
	Aarseth, S. J., \& Fall, S. M. 1980, \apj, 236, 43
\bibitem[2004]{gdds}
        Abraham, R. G., Glazebrook, K., McCarthy, P. J., \etal 2004, \aj\, in press, (astro-ph/0402436)
\bibitem[1993]{aragon}
	Arag\'{o}n-Salamanca, A., Ellis, R. S., Couch, W. J., \& Carter, D. 1993, \mnras, 262, 764 (A93)
\bibitem[2002]{bahcall02}
	Bahcall, N. A., \& Bode, P. 2003, \apjl, 588, 1
\bibitem[1999]{benitez}
	Ben\'{\i}tez, N., Broadhurst. T., Rosati, P., \etal 1999, \apj, 527, 31
\bibitem[1996]{bertin}
	Bertin, E., \& Arnouts, S. 1996, \aaps, 117, 393
\bibitem[2000]{best}
        Best, J. S. 2000, \apj, 541, 519
\bibitem[2003]{blakeslee}
        Blakeslee, J. P., Franx, M., Postman, M., \etal 2003, \apjl, 596, 143 
\bibitem[1999]{blanton99}
	Blanton, M., Cen, R., Ostriker, J. P., \& Strauss, M. A. 1999, \apj, 522, 590
\bibitem[2000]{blanton00}
	Blanton, M., Cen, R., Ostriker, J. P., Strauss, M. A., \& Tegmark, M. 2000, \apj, 531, 1
\bibitem[1984]{blumenthal}
        Blumenthal, G.R., Faber, S. M., Primack, J. R., \& Rees, M. J. 1984, \nat, 311, 517
\bibitem[2000]{bolzonella}
	Bolzonella, M., Miralles, J.-M., \& Pell\'{o}, R. 2000, \aap, 363, 476
\bibitem[1992]{bower}
	Bower, R. G., Lucey, J. R., \& Ellis, R. S. 1992, \mnras, 254, 601
\bibitem[2003]{brand}
        Brand, K., Rawlings, S., Hill, G. J., Lacy, M., Mitchell, E., \& Tufts, J. 2003, \mnras, 344, 283
\bibitem[1993]{bruzual}
	Bruzual, G., \& Charlot, S. 1993, \apj, 405, 538
\bibitem[1984]{butcher}
	Butcher, H., \& Oemler, A. 1984, \apj, 285, 426
\bibitem[2000]{calzetti}
	Calzetti, D., Armus, L., Bohlin, R. C., \etal 2000, \apj, 533, 682
\bibitem[1991]{clowes91}
	Clowes, R. G., \& Campusano, L. E. 1991, \mnras, 249, 218
\bibitem[1994]{clowes94}
	Clowes, R. G., \& Campusano, L. E. 1994, \mnras, 266, 317
\bibitem[1999]{clowes99}
	Clowes, R. G., Campusano, L. E., \& Graham, M. J. 1999, \mnras, 309, 48
\bibitem[2001]{colless}
        Colless, M., Dalton, G., Maddox, S., \etal 2001, \mnras, 328, 1039
\bibitem[1987]{crampton87}
	Crampton, D., Cowley, A. P., \& Hartwick, F. D. A. 1987, \apj, 314, 129
\bibitem[1989]{crampton89}
	Crampton, D., Cowley, A. P., \& Hartwick, F. D. A. 1989, \apj, 345, 59
\bibitem[2001]{croom}
	Croom, S. M., Shanks, T., Boyle, B. J., \etal 2001, \mnras, 325, 483
\bibitem[1999a]{dd99a}
	Demianski, M., \& Doroshkevich, A. G. 1999a, \apj, 512, 527
\bibitem[1999b]{dd99b}
	Demianski, M., \& Doroshkevich, A. G. 1999b, \mnras, 306, 779
\bibitem[1999]{depropris}
	de Propris, R., Stanford, S. A., Eisenhardt, P. R., Dickinson, M., \& Elston, R. 1999, \aj, 118, 719
\bibitem[1996]{d96}
	Doroshkevich, A. G., Tucker, D. L., Oemler, A., \etal 1996, \mnras, 283, 1281
\bibitem[1999]{d99}
	Doroshkevich, A. G., M\"{u}ller, V., Retzlaff, J., \& Turchaninov, V. L. 1999, \mnras, 306, 575
\bibitem[2000]{d00}
	Doroshkevich, A. G., Fong, R., McCracken, H. J., \etal 2000, \mnras, 315, 767
\bibitem[2001]{d01}
	Doroshkevich, A. G., Tucker, D. L., Fong, R., Turchaninov, V. L., \& Lin, H. 2001, \mnras, 322, 369
\bibitem[2002]{d02}
	Doroshkevich, A. G., Tucker, D. L., \& Allam, S. 2002, (astro-ph/0206301)
\bibitem[1997]{ellis}
	Ellis, R. S., Smail, I., Dressler, A., \etal 1997, \apj, 483, 582
\bibitem[2003]{enoki}
        Enoki, M., Nagashima, M., \& Gouda, N. 2003, \pasj, 55, 133
\bibitem[2002]{evrard}
	Evrard, A. E., MacFarland, T. J., Couchman, H. M., \etal 2002, \apj, 573, 7
\bibitem[1989]{geller}
	Geller, M. J., \& Huchra, J. P. 1989, Sci, 246, 897
\bibitem[2000]{gladders}
	Gladders, M. D., \& Yee, H. K. C. 2000, \aj, 120, 2148
\bibitem[1995]{graham95}
	Graham, M. J., Clowes, R. G., \& Campusano, L. E. 1995, \mnras, 275, 790
\bibitem[1993]{haehnelt}
	Haehnelt, M. G., \& Rees, M. J. 1993, \mnras, 263, 168
\bibitem[2001]{hainesphd}
	Haines, C. P. 2001, PhD thesis, Univ. of Central Lancashire
\bibitem[2001a]{haines01b}
	Haines, C. P., Clowes, R. G., \& Campusano, L. E. 2001a, The New Era of Wide Field Astronomy, eds. R. Clowes, A. Adamson, G. Bromage, ASP Conf. Ser. vol. 232, p. 117
\bibitem[2001b]{haines01a}
	Haines, C. P., Clowes, R. G., Campusano, L. E., \& Adamson, A. J. 2001b, \mnras, 323, 688
\bibitem[2002]{hoyle}
	Hoyle, F., Outram, P. J., Shanks, T., \etal 2002, \mnras, 329, 336
\bibitem[1995]{hutchings}
	Hutchings, J. B., Crampton, D., \& Johnson, A., 1995, \aj, 109, 73
\bibitem[1984]{kaiser}
        Kaiser, N. 1984, \apjl, 284, 9
\bibitem[1997]{ka97}
	Kodama, T., \& Arimoto, N. 1997, \aap, 320, 41
\bibitem[1998]{kodama}
	Kodama, T., Arimoto, N., Barger, A. J., \& Arag\'{o}n-Salamanca, A. 1998, \aap, 334, 99
\bibitem[1994]{komberg}
	Komberg, B. V., \& Lukash, V. N. 1994, \mnras, 269, 277
\bibitem[2001]{kukula}
	Kukula, M. J., Dunlop, J. S., McLure, R. J., \etal 2001, \mnras, 326, 1533
\bibitem[1992]{landolt}
	Landolt, A. U. 1992, \aj, 104, 340
\bibitem[1991]{longo}
	Longo, M. J. 1991, \apjl, 372, 59
\bibitem[2002]{madgwick}
	Madgwick, D. S., Lahav, O., Baldry, I. K., \etal 2002, \mnras, 333, 133
\bibitem[2001]{mccarthy}
	McCarthy, P. J., Carlberg, R. G., Chen, H.-W., \etal 2001, \apjl, 560, 131
\bibitem[1999]{mclure99}
	McLure, R. J., Kukula, M. J., Dunlop, J. S., \etal 1999, \mnras, 308, 377
\bibitem[2001]{merritt}
	Merritt, D., \& Ferrarese, L. 2001, \mnras, 320, L30
\bibitem[2001]{mb}
        Miller, C. J., \& Batuski, D. J. 2001, \apj, 551, 635
\bibitem[1979]{miller}
	Miller, G. E., \& Scalo, J. M. 1979, \apjs, 41, 513
\bibitem[2004]{miller04}
        Miller, L., Croom S. M., Boyle, B. J., \etal 2004, \mnras\, submitted (astro-ph/0403065)
\bibitem[2001]{nelson}
	Nelson, A. E., Gonzales, A. H., Zaritsky, D., \& Dalcanton, J. J. 2001, \aj, 563, 629 (N01)
\bibitem[1999]{newman99}
	Newman, P. R. 1999, PhD thesis, Univ. of Central Lancashire
\bibitem[1998]{newman98}
	Newman, P. R., Clowes, R. G., Campusano, L. E., \& Graham, M. J. 1998, Wide Field Surveys in Cosmology, 14th IAP meeting, \'{E}ditions Frontieres, p. 408
\bibitem[2003]{ohta}
	Ohta, K., Akiyama, M., Ueda, Y., \etal 2003, \apj, 598, 210
\bibitem[1998]{persson}
	Persson, S. E., Murphy, D. C., Krzeminski, W., Roth, M., \& Rieke, M., J 1998, \aj, 116, 2475
\bibitem[1993]{pisani93}
	Pisani, A. 1993, \mnras, 265, 706
\bibitem[1996]{pisani96}
	Pisani, A. 1996, \mnras, 278, 697
\bibitem[1992]{ramella}
	Ramella, M., Geller, M. J., \& Huchra, J. P. 1992, \apj, 384, 396
\bibitem[1999]{rosati}
	Rosati, P., Stanford, S. A., Eisenhardt, P. R., \etal 1999, \aj, 118, 76
\bibitem[1999]{sanchez99}
	S\'{a}nchez, S. F., \& Gonz\'{a}lez-Serrano 1999, \aap, 352, 383
\bibitem[2002]{sanchez02}
	S\'{a}nchez, S. F., \& Gonz\'{a}lez-Serrano 2002, \aap, 396, 773
\bibitem[1976]{schechter}
	Schechter, P. 1976, \apj, 304, 297
\bibitem[1996]{shectman}
	Shectman, S. A., Landy, S. D., Oemler, A., \etal 1996, \apj, 472, 170
\bibitem[1997]{smail97}
	Smail, I., Dressler, A., Couch, W. J., \etal 1997, \apjs, 110, 213
\bibitem[2002]{sochting}
	S\"{o}chting, I. K., Clowes, R. G., \& Campusano, L. E. 2002, \mnras, 331, 569
\bibitem[2003]{sochtingphd}
	S\"{o}chting, I. K. 2003, PhD thesis, Univ. of Central Lancashire
\bibitem[2003]{spergel}
        Spergel, D. N., Verde, L., Peiris, H. V., \etal 2003, \apjs, 148, 175
\bibitem[1998]{stanford98}
	Stanford, S. A., Eisenhardt, P. R., \& Dickinson, M. 1998, \apj, 492, 461 (SED98)
\bibitem[1997]{stanford97}
	Stanford, S. A., Elston, R., Eisenhardt, P. R., \etal 1997, \aj, 114, 2232 (S97)
\bibitem[2002]{stanford02}
	Stanford, S. A., Holden, B., Rosati, P., \etal 2002, \aj, 123, 619
\bibitem[2000]{tanaka00}
	Tanaka, I., Yamada, T., Arag\'{o}n-Salamanca, A., \etal 2000, \apj, 528, 123
\bibitem[2001]{tanaka01}
	Tanaka, I., Yamada, T., Turner, E. L., \& Suto, Y. 2001, \apj, 547, 521
\bibitem[1982]{webster}
	Webster, A. 1982, \mnras, 199, 683
\bibitem[2002]{williger}
	Williger, G. M., Campusano, L. E., Clowes, R. G., \& Graham, M. J. 2002, \apj, 578, 708
\bibitem[2000]{wold00}
        Wold, M., Lacy, M., Lilje, P. B., \& Serjeant, S. 2001, \mnras, 316, 267
\bibitem[2001]{wold01}
        Wold, M., Lacy, M., Lilje, P. B., \& Serjeant, S. 2001, \mnras, 323, 231
\bibitem[2002]{wittman}
	Wittman, D., Tyson, J. A., Dell'Antonio, I. P., \etal 2002, \procspie, vol. 4836, 73
\bibitem[2000]{york}
        York, D. G., Adelman, J., Anderson, J. E., \etal 2000, \aj, 120, 1579
\end{thebibliography}
\end{document}